\documentclass[10pt]{article}
\pdfoutput=1
\usepackage{jheppub,amsmath,amssymb}
\usepackage{enumerate}
\usepackage{bm}
\usepackage{bbm}
\usepackage{enumitem}
\usepackage[usenames,dvipsnames]{xcolor}
\usepackage{amsmath}
\usepackage{mathtools}
\usepackage{amsfonts}
\usepackage{amssymb}
\usepackage{graphicx}
\usepackage{hhline}
\usepackage{hyperref}
\usepackage{color}
\usepackage{verbatim}
\usepackage{amsmath,amsthm,amssymb}
\usepackage{lscape}
\usepackage{float}
\usepackage{caption}
\usepackage{lscape}
\usepackage{breqn}
\usepackage{multicol}
\usepackage{graphics}
\usepackage{tikz,todonotes}
\usepackage[utf8]{inputenc}
\usepackage{autobreak}
\usetikzlibrary{arrows,positioning,decorations.markings,decorations.pathmorphing,calc}
\pgfdeclarelayer{edgelayer}
\pgfdeclarelayer{nodelayer}
\usetikzlibrary{decorations.pathreplacing}
\pgfsetlayers{edgelayer,nodelayer,main}
\tikzset{none/.style={draw=none}}
\tikzset{new edge style 2/.style={black}}
\tikzset{new style 0/.style={black}}
\tikzset{rednode/.style={draw=none, scale=0.3pt,fill=red,circle, draw}}
\tikzset{redline/.style={line width=0.3mm,red}}
\tikzset{greyE/.style={line width=0.1mm,gray}}
\usepackage[utf8]{inputenc}
\usetikzlibrary{arrows,positioning,decorations.markings,decorations.pathmorphing,calc}
\usepackage{tikz-feynman}
\usepackage{subcaption}

\definecolor{hyperref}{RGB}{026,028,087}

\newcommand{\A}{{\cal A}}

\newcommand{\beq}{\begin{equation}}
\newcommand{\eeq}{\end{equation}}
\newcommand{\bea}{\begin{eqnarray}}
\newcommand{\eea}{\end{eqnarray}}
\def\be{\begin{equation}}
\def\ee{\end{equation}}

\def\beq{\begin{equation}}
\def\eeq{\end{equation}}

\newcommand{\mpl}{M_{\rm Pl}}

\renewcommand{\[}{\left[}
\renewcommand{\]}{\right]}
\renewcommand{\L}{\mathcal L}

\renewcommand{\Im}{{\rm Im}}

\def\be{\begin{equation}}
\def\ee{\end{equation}}
\def\ba{\begin{eqnarray}}
\def\ea{\end{eqnarray}}

\def\nn{\nonumber}

\def\d{\mathrm{d}}

\newcommand{\msbar}{\overline{\text{MS}}}

\usepackage[normalem]{ulem}

\def\ba{\begin{eqnarray}}
\def\ea{\end{eqnarray}}

\def\L{\mathcal{L}}

\def\d{\mathrm{d}}

\def\({\left(}
\def\){\right)}

\def\mpl{M_{\rm Pl}}
\def\p{\partial}

\begin{document}

\hfill Imperial/TP/2020/LA/02\\

\title{Positivity Bounds and the Massless Spin-2 Pole}

\author[a]{Lasma Alberte,}
\author[a,b]{Claudia de Rham,}
\author[a]{Sumer Jaitly,}
\author[a,b]{Andrew J. Tolley}
\affiliation[a]{Theoretical Physics, Blackett Laboratory, Imperial College, London, SW7 2AZ, U.K.}
\affiliation[b]{CERCA, Department of Physics, Case Western Reserve University, 10900 Euclid Ave, Cleveland, OH 44106, USA}

\emailAdd{l.alberte@imperial.ac.uk}
\emailAdd{c.de-rham@imperial.ac.uk}
\emailAdd{sumer.jaitly14@imperial.ac.uk}
\emailAdd{a.tolley@imperial.ac.uk}

\abstract{The presence of a massless spin-2 field in an effective field theory results in a $t$-channel pole in the scattering amplitudes that precludes the application of standard positivity bounds. Despite this, recent arguments based on compactification to three dimensions have suggested that positivity bounds may be applied to the $t$-channel pole subtracted amplitude. If correct this would have deep implications for UV physics and the Weak Gravity Conjecture. Within the context of a simple renormalizable field theory coupled to gravity we find that applying these arguments would constrain the low-energy coupling constants in a way which is incompatible with their actual values. This contradiction persists on deforming the theory. Further enforcing the $t$-channel pole subtracted positivity bounds on such generic renormalizable effective theories coupled to gravity would imply new physics at a scale parametrically smaller than expected, with far reaching implications.
This suggests that generically the standard positivity bounds are inapplicable with gravity and we highlight a number of issues that impinge on the formulation of a three-dimensional amplitude which simultaneously satisfies the required properties of analyticity, positivity and crossing symmetry. We conjecture instead a modified bound that ought to be satisfied independently of the precise details of the high energy completion. }

\maketitle


\section{Introduction}
Over the past few decades effective field theories (EFTs) have proven to be an incredibly powerful tool for studying physical systems at both high and low energies, with applications in all areas of physics ranging from particle physics to cosmology and condensed matter. While it is almost always possible to come up with an EFT valid in a given energy range that correctly describes the physical problem in question, a theoretically more compelling question is whether a given low-energy EFT can be successfully UV completed into another theory valid at higher energies. The answer depends strongly on what requirements one wishes to impose on the high energy theory. Requiring that the low-energy EFT has a \emph{standard} UV completion that is Lorentz invariant, local and causal is known to impose strong constraints on the coefficients in the low-energy action \cite{Pham:1985cr,Ananthanarayan:1994hf,Adams:2006sv}. These are known as the positivity bounds and can be imposed on the scattering amplitudes in the low-energy theory by using the axioms of the S-matrix theory, mostly relying on its analyticity properties. First developed for scalar field theories in the forward limit, the positivity bounds have been since generalized to particles with spin \cite{Cheung:2016yqr,Bonifacio:2016wcb,Bellazzini:2016xrt,deRham:2017zjm,deRham:2017imi,deRham:2017xox,deRham:2018qqo,Kim:2019wjo,Alberte:2019xfh,Alberte:2019zhd,Remmen:2019cyz} and extended away from the forward limit \cite{deRham:2017avq}. Including any additional known information about the low-energy EFT (e.g. calculable low-energy loop diagrams) has enabled to further expand the applications of the positivity bounds, going under the name of \emph{improved positivity bounds} \cite{deRham:2017xox,deRham:2017imi,Bellazzini:2016xrt}.\\

From the point of view of practical applications of positivity bounds to real world EFTs, one of the assumptions that turns out to be the most restrictive is that of polynomial boundedness of the scattering amplitudes in the complex $s$-plane, inferred from locality. The famous Froissart bound \cite{Froissart:1961ux,Martin:1965jj}, extended beyond the forward limit in \cite{PhysRev.135.B1375}, states that in the presence of a mass gap, any local $2-2$ scattering amplitude should not grow faster than the fourth power of the center of mass energy at sufficiently high energy.  Technically speaking, the requirement of the existence of a mass gap makes the positivity bounds not directly applicable to one of the most intriguing low-energy effective theories --- general relativity (GR). Nevertheless, it is typically expected, even with gravity, that the scattering amplitude will be bounded at least in the Jin-Martin \cite{PhysRev.135.B1375} sense $\lim_{|s|\to \infty}s^{-2}\mathcal A(s,t)=0$ at fixed $t$, as is argued to be the case in string theory for fixed momentum transfer scattering \cite{Giddings:2009gj}, despite the violation of polynomial boundedness. This justifies attempting to apply the positivity bounds to gravitational EFTs \cite{Cheung:2016wjt,Bellazzini:2015cra}. \\

A more serious issue associated with applying positivity bounds to gravitational theories, is the presence of an infamous $t$--channel pole whose residue grows faster than the Froissart or Jin-Martin bound. This growth implies that the pole cannot be subtracted, or else the resulting subtracted amplitude would itself violate the Froissart bound. More precisely, this would contradict the assumption that we can write a dispersion relation for the pole subtracted scattering amplitude with only two subtractions. More importantly though, the pole at $t=0$, and the associated branch point that arises from graviton loops, prevents the analytic continuation of the partial wave expansion from the physical region $t<0$ to $t \ge 0$ which is a crucial step in deriving positivity bounds. As a result it is impossible to use the positivity bounds in their standard form for processes that exhibit a spin-$J$ pole with $J\ge2$. See Ref.~\cite{Tokuda:2020mlf} for a related recent discussion which parallels some of our findings. \\

Recently, a novel way to deal with this $t$-channel pole was suggested in \cite{Bellazzini:2019xts}, which relies on compactifying one of the spatial dimensions on a circle.  Since there are no propagating massless gravitons in three dimensions (3d) one would expect the unpleasant massless spin-2 $t$-channel pole\footnote{In addition, when graviton loops are included there is a branch cut which extends to $t=0$. The two effects come in at a different order in $1/\mpl^2$ which allows us to cleanly separate them. } to be absent from the scattering amplitude. While in fact the term is still present perturbatively in the amplitude written in Mandelstam variables, there is indeed no physical massless graviton mediating the exchange.
Furthermore, in Mandelstam variables the problematic part of the residue is removed after resumming the contributions from higher order ladder diagrams (see, e.g. \cite{Ciafaloni:1992hu} and appendix~A of \cite{Bellazzini:2019xts}). Motivated by this observation, one can then argue that the standard positivity bound should apply to 3d scattering amplitudes. This technique would then allow us to  constrain coefficients in the low-energy EFTs which were previously beyond the reach of the positivity bounds programme in four spacetime dimensions (4d). Interestingly, from the 4d point of view these bounds are equivalent  to what one would have deduced, would the $t$-channel pole be simply disregarded. If applicable, this would be a remarkable result with far--reaching implications. It could potentially open a whole new window on investigating the higher derivative corrections in either GR itself or in any theory that includes massless spin-2 or higher fields. We will refer to these as `{\it compactified positivity bounds}'.\\

 In \cite{Bellazzini:2019xts} the implications of these compactified positivity bounds on the higher derivative corrections to the Einstein--Maxwell theory were studied in relation to the weak gravity conjecture (WGC) \cite{ArkaniHamed:2006dz}. It was found that the positivity bounds imply that extremal black holes of mass $M$ and charge $Q$ must satisfy $\mpl Q/M> 1$, exceeding the extremal charge-to-mass ratio in GR and thus proving one of the versions of the weak gravity conjecture as suggested in \cite{Cheung:2014ega,Hamada:2018dde}. Another remarkable example of possible consequences of these positivity bounds was studied in \cite{deRham:2019ctd} where the impact of the backreaction of matter fields on the propagation speed of the graviton was studied. It was found that imposing the compactified positivity bounds on the couplings of the higher curvature terms arising after integrating out matter fields leads to apparent superluminal propagation speed for gravity on cosmological backgrounds. Nevertheless there remains no violation of causality \cite{deRham:2020zyh,Hollowood:2015elj}.\\

As the previous few examples already show, applications of these compactified positivity bounds to gravitational theories might provide genuine insights in our understanding of gravity at high energy. Nevertheless, a few open questions remain on the validity of the procedure. Besides the subtleties associated with defining the massless asymptotic states  in 3d, it may also be unclear whether  the forward limit necessarily commutes with the `de-compactification' limit (where the size of the compactified circle is sent to infinity) \cite{Loges:2019jzs}. So far, in most cases where the compactified positivity bounds have been applied, neither a full nor a partial UV completion of the low-energy theories analyzed was in fact known and hence there was no explicit way of testing their predictions. Motivated by the current state-of-affairs, in this work we shall therefore apply the compactified positivity bounds to an IR theory for which a partial UV completion\footnote{By partial UV completion we mean an effective theory valid at a higher energy scale than the original low-energy EFT. Our partial UV completions shall be theories which are renormalizable in the absence of gravity and for which the traditional expectation would be that the cutoff of the EFT is $\Lambda \sim \mpl$.} is in fact known, hence providing an explicit framework where the validity of the compactified positivity bounds can be put to the test. \\

In the absence of gravity, positivity bounds are automatically satisfied for renormalizable field theories. If they were not, it would be necessary to include irrelevant operators to satisfy them, contradicting the assumption of renormalizability. Our central interest will be renormalizable theories coupled to gravity. By construction, any terms that may potentially violate positivity will necessarily vanish in the decoupling limit $\mpl \rightarrow \infty$.
A fundamentally important example of a renormalizable theory is none other than QED. Upon minimally coupling QED to gravity, renormalizability is of course spoiled, but this spoiling is suppressed by the Planck scale. Naively, one would expect QED or a generic renormalizable theory minimally coupled to gravity to be a
perfectly well-defined partial UV completion. Moreover from decoupling arguments, at sufficiently low-energy scales we would expect physics to be insensitive to how QED is fundamentally implemented into a full UV complete quantum theory of gravity.\\

Inspired by QED, in this work we consider a scalar photon QED toy model where both the vector field and the electron are treated as scalar fields \cite{Hollowood:2015elj} while leaving the case of the actual QED to a separate work~\cite{newpaper}. As we shall see, the compactified positivity bounds that we would have derived from compactifying one dimension are inconsistent with the knowledge  obtained directly from the (partial) UV theory. This represents an explicit example where the positivity bounds inferred from 3d (or 4d compactified on a circle) would have led to the wrong conclusions, hence casting doubt on the generic validity of the procedure. Naturally, the example we provide has its own limitations:
\begin{itemize}
\item First the model we shall propose only has a known {\it partial} UV completion and not a full one, its non-renormalizability arises entirely from graviton exchange/loops. Nonetheless, this limitation is a weak one as corrections from UV physics will be suppressed by additional powers of $\mpl$. Although one may argue that our particular example renormalizable field theory belongs to the `swampland', we will show that these features are in fact common to generic renormalizable field theories including QED itself \cite{newpaper}, and hence applicable to theories which are known to arise from consistent UV completions.

\item Second, and perhaps more to the point, the partial UV completion we are dealing with is not a tree-level completion (for example of the string/Regge higher spin type \cite{Hamada:2018dde})\footnote{Often in the literature this is referred to as a weakly coupled UV completion on the grounds that there must be some dimensionless parameter that suppressed loops, a role played by the string coupling constant $g_s$ (dilaton) in string theory. However this terminology can be confusing since the completion we consider is itself weakly coupled in the sense that perturbation theory is under control, but the key contribution to the scattering amplitudes arise at one-loop in the heavy field.} since the relevant effect from the electron arises at one loop.  Yet the beauty of the positivity bounds is that they are supposed to be agnostic of the precise type of UV completions one is dealing with, so long as it is local, Lorentz-invariant, unitary and causal. The argument of \cite{Bellazzini:2019xts} if valid, should apply equally well for these heavy loop completions.
 \end{itemize}

The derivation of the compactified positivity bounds rests on the applicability of positivity bounds to 3d gravitational scattering amplitudes. However these are notoriously poorly defined. We critically assess this in section~\ref{Infrared} where we note that in 3d there is no scattering amplitude which simultaneously satisfies: {(a) analyticity and finiteness in the forward scattering limit, (b) positivity of its imaginary part and (c) crossing symmetry}. Since these are crucial assumptions in the derivation of positivity bounds, this casts a significant doubt on the applicability of positivity bounds with (massless) gravity. Interestingly these issues disappear when considering the exchange of massive spin-2 fields.
On the other hand, for a massless exchange, we show that the issue with the $t$-channel pole in 4d ends up manifesting itself through a slightly different but ultimately equivalent violation of analyticity in 3d, more precisely through the presence of a delta function. Attempting to applying the positivity bounds to 3d amplitudes where the delta function is removed is ultimately not justified and can lead to incorrect implications. \\

The rest of this paper is organized as follows. In section~\ref{sec:outline} we review the various positivity bounds. We then  introduce the scalar photon QED in section~\ref{sec:qed}, and present the core of the inconsistency that one runs into when applying the compactified positivity bounds of \cite{Bellazzini:2019xts} to the low-energy EFT. In section~\ref{sec:uv}
we refine this discussion by introducing a spectator field to probe the consistency of the partial UV completion, and we summarize the calculation of the amplitude in the low-energy effective theory and the partial UV completion both before and after compactification. In section~\ref{sec:newUV} we address the question of whether the failing lies in our choice of partial UV completion by amending it with both renormalizable and irrelevant operators indicative of new UV physics. We find that only new physics at a parametrically low scale $\Lambda \sim (M \mpl)^{1/2}$  can ensure the compactified positivity bounds are satisfied {and emphasize the far-reaching implications of these arguments, particularly when applied to other fields like dark matter, which would not have otherwise been expected to couple directly to Standard Model fields at such a low scale}. In section~\ref{Infrared} we take the opposite perspective and address possible flaws with the derivation of \cite{Bellazzini:2019xts}, principally those due to the ill-defined nature of 3d scattering amplitudes with massless spin-2 exchange. We point out that  positivity bounds are cleanly respected for massive spin-2 states. However for massless spin-2 states, the issue with the 4d $t$-channel pole manifests itself through a delta function in 3d which ultimately precludes the application of the positivity bounds in 3d, just as it was in 4d. We conclude in section~\ref{sec:conclusions} by conjecturing a bound that ought to be satisfied even in the presence of a t-channel pole, that requires no further assumptions or limitation on the UV behaviour other than the standard Unitarity, Locality, Lorentz Invariance and Causality requirements.
All further details and consistency checks are presented in appendices. In particular in appendix~\ref{sec:3d} we discuss the compactification procedure applied to our particular partial UV completion.

\section{Positivity Bounds --- the relevance of the $t$--channel pole}\label{sec:outline}

In this section we lay out the procedure by which we check the consistency and implications of the compactified positivity bounds of \cite{Bellazzini:2019xts}. We shall start from a known partial UV theory containing a heavy field that can be integrated out leading to a known IR theory. Having both the partial UV and the IR theory at hand allows us to directly check whether the constraints that the new positivity bounds impose on the couplings in the low-energy theory are satisfied by the information that we have from the partial UV theory. As our UV theory we shall use a QED-type scalar field theory with a massive `electron' that we shall integrate out to obtain the IR theory --- a scalar analogue of the Einstein--Maxwell theory with higher derivative corrections with specific calculable couplings on which we then impose the new positivity bounds as described below.

\subsection{Positivity bounds}\label{sec:positivity}
Let us start with a short review of the positivity bounds. The standard positivity bounds (see e.g. \cite{Pham:1985cr,Ananthanarayan:1994hf,Adams:2006sv,Cheung:2016yqr,Bellazzini:2016xrt,deRham:2017avq}) can be applied to theories where all the fields have a mass gap, or are regulated by adding a small mass that can later be sent to zero. We consider the elastic $2-2$ scattering  ($A+B \rightarrow A+B$) of particles of mass $m_1$ and $m_2$, which necessarily respects $s-u$ crossing symmetry, denoting by $s, t$ and $u=2m_1^2+2m_2^2-s-t$ the standard Mandelstam variables. The analyticity properties of the elastic scattering amplitude $\mathcal A(s,t)$  allows to express it as a dispersion relation:
\be\label{Cauchy}
\mathcal A(s,t)=\frac{1}{2\pi i}\oint_{\mathcal C} \text{d} s'\frac{\mathcal A(s',t)}{(s'-s)}\,,
\ee
where $\mathcal C$ is a contour in the complex $s'$ plane that includes the point $s$ and excludes poles and branch cuts.

The essence of the positivity bounds relies on the optical theorem for elastic scatterings with the same particle content, $i$, in the initial and final state of the scattering. The imaginary part of the forward limit amplitude of such scattering process is then
\be
\label{eq:optical1}
\text{Im}\,\mathcal A_i(s,0)=\frac{1}{2}\sum_f\int\text{d}\Pi_f |\mathcal A_{i\to f}|^2\,,
\ee
where $\mathcal A_{i\to f}$ is the scattering amplitude for the process $i\to f$ with $f$ denoting all possible intermediate states and $\text {d}\Pi_f$ stands for the phase volume. Importantly, this tells us that the contribution from \emph{each} of the possible intermediate scattering processes gives a positive contribution to the imaginary part of the amplitude. In terms of the $2-2$ elastic scattering amplitude $\mathcal A$, this tells us that within the physical region,  $\text{Im}\,\mathcal A(s,0)>0$. Further positivity of the imaginary part of individual partial waves $\text{Im}\,a_l(s) >0$, taken together with the partial wave expansion, implies $\partial_t^N \text{Im}\,\mathcal A(s,0)>0$ for all integer $N \ge 0$. Making use of the Jin--Martin~\cite{PhysRev.135.B1375} extension of the Froissart bound $\lim_{|s|\to \infty}s^{-2}\mathcal A(s,t)=0$ at fixed $t$, and $s-u$ crossing symmetry, we infer from \eqref{Cauchy} a dispersion relation for the amplitude with two subtractions
\ba\label{dispersion}
\mathcal A(s,t) &=& a_0(t) + a_1(t) s +  \sum_I \frac{\lambda_I(t)}{m_I^2-s} + \sum_I \frac{\lambda_I(t)}{m_I^2-u} \nn \\
&+& \frac{s^2}{\pi}\int_{(m_1+m_2)^2}^\infty\text{d}s'\,\frac{\text{Im}\,\mathcal A(s',t)}{{s'}^2 (s'-s)}+\frac{u^2}{\pi}\int_{(m_1+m_2)^2}^\infty\text{d}s'\,\frac{\text{Im}\,\mathcal A(s',t)}{{s'}^2 (s'-u)} \, ,
\ea
where we see clearly the separation between the poles at $s=m_I^2$, $u=m_I^2$ and the left-hand and right-hand branch cuts.
Defining $\tilde{\mathcal A}(s,t)$ as the pole-subtracted amplitude
\be
\tilde{\mathcal A}(s,t)\equiv \mathcal A(s,t) -\sum_I \frac{\lambda_I(t)}{m_I^2-s} - \sum_I \frac{\lambda_I(t)}{m_I^2-u} \, ,
\ee
we easily infer the positivity bounds:
\be\label{positivity10}
\frac{\text{d}^2\tilde{\mathcal A}(s,t)}{\text{d}s^2}>0\,,
\ee
valid for  $(m_1-m_2)^2 -t <s < (m_1+m_2)^2$ together with $ 0 \le t < {\rm Min}(4m_1^2,4m_2^2,m_J^2)$\footnote{The precise range of $t$ depends on what processes are allowed by symmetries and kinematics. It may for instance be the higher value $0 \le t < (m_1+m_2)^2$.} where $m_J^2$ is the smallest mass of particles with spin $J>2$ in the spectrum. For our present purposes it is sufficient to utilize these in the forward limit $t=0$
\be\label{positivity}
\frac{\text{d}^2\tilde{\mathcal A}(s,0)}{\text{d}s^2}>0\,,  \quad (m_1-m_2)^2  <s < (m_1+m_2)^2 \, .
\ee
Additional extensions beyond the forward limit are given for example in \cite{deRham:2017avq}.
Note that $\tilde{\mathcal A}(s,t)$ denotes the $s$-channel and $u$-channel pole-subtracted amplitude. Ideally we would also like to remove the $t$-channel pole. Since the residue of a $t$-channel pole associated with the exchange of a spin $J$ particle scales as $s^J/t$ at large $s$, we can only subtract the $t$-channel pole for exchanged states with spin $J<2$ without contradicting the assumption that the dispersion relation \eqref{dispersion} only has two subtractions. This brings us to the essential point, in a gravitational theory with a massless graviton, the $t$--channel pole scales as $s^2/t$ as in \eqref{amp_pUV} and so cannot be removed without contradicting the assumption of two subtractions. Therefore, in what follows the amplitude  $\tilde{\mathcal A}(s,t)$ shall generically denote the amplitude for which only the $s$ and $u$--channel poles of the massless graviton exchange have been removed but keeping the $t$--channel pole, whereas the $s$, $t$ and $u$ channel poles will be removed for particles with spin $J<2$. Since the $t$-pole cannot be subtract it dominates the forward limit rendering \eqref{positivity} contentless.\\

If a massless spin $J \ge 2$ pole is present, there is a second major problem with the standard arguments. The proof that  $\partial_t^N \text{Im}\,\mathcal A(s,0)>0$ assumes that the partial wave expansion can be continued from the physical region $t \le 0$ to the unphysical region $t >0$ at least in the neighbourhood of $t=0$. The presence of a pole and indeed the branch point from graviton loops undermines this. This problem is conveniently avoided for $J<2$ by analytically continuing the partial wave expansion for $\partial^2_s\tilde{\mathcal A}(s,t)$ for which the associated pole drops out. There remains the branch cut, but these contributions are typically suppressed by a loop counting parameter. For $J=2$ we can continue the partial wave expansion for $\partial^3_s\tilde{\mathcal A}(s,t)$, at least provided we neglect graviton loops, but this amounts to performing one additional subtraction in the dispersion relation~\eqref{dispersion}:
\ba\label{dispersion2}
\mathcal A(s,t) &=& a_0(t) + a_1(t) s + a_2(t) s^2+ \sum_I \frac{\lambda_I(t)}{m_I^2-s} + \sum_I \frac{\lambda_I(t)}{m_I^2-u}  \\
&+& \frac{s^3}{\pi}\int_{(m_1+m_2)^2}^\infty\text{d}s'\,\frac{\text{Im}\,\mathcal A(s',t)}{{s'}^3 (s'-s)}+\frac{u^3}{\pi}\int_{(m_1+m_2)^2}^\infty\text{d}s'\,\frac{\text{Im}\,\mathcal A(s',t)}{{s'}^3 (s'-u)} \, ,\nn
\ea
with the $J=2$ $t$-channel pole now contained in $a_2(t)$. While we can infer positivity statements for higher order $s$-derivative positivity bounds, we lose the most valuable information, namely the condition \eqref{positivity}, since $a_2(t)$ cannot be determined by analyticity alone.

\subsection{Positivity of Spectral Flow}\label{Spectral flow}

When a massless spin $J = 2$ pole is present we cannot immediately prove positivity of the $t$-pole subtracted coefficient $a_2(t)$ for the reasons discussed, however we can prove  positivity of its spectral flow. Following a similar discussion in \cite{deRham:2020zyh}, we may rewrite the dispersion relation \eqref{dispersion2} by performing the subtractions at an arbitrary scale $s'=-\mu_0$ to give
\ba\label{dispersion3}
\mathcal A(s,t) &=& b_0(\mu_0,t) + b_2(\mu_0,t) ((s+\mu_0)^2+(u+\mu_0)^2)+ \sum_I \frac{\lambda_I(t)}{m_I^2-s} + \sum_I \frac{\lambda_I(t)}{m_I^2-u}  \\
&+& \frac{(s+\mu_0)^3}{\pi}\int_{(m_1+m_2)^2}^\infty\text{d}s'\,\frac{\text{Im}\,\mathcal A(s',t)}{{(s'+\mu_0)}^3 (s'-s)}+\frac{(u+\mu_0)^3}{\pi}\int_{(m_1+m_2)^2}^\infty\text{d}s'\,\frac{\text{Im}\,\mathcal A(s',t)}{{(s'+\mu_0)}^3 (s'-u)} \, .\nn
\ea
The subtraction constant $ b_2(\mu_0,t) $ is now a function of the chosen energy scale. However, since the amplitude cannot depend on $\mu_0$ we may easily derive the `renormalization group' style equation
\be
\mu_0 \frac{\partial b_2(\mu_0,t) }{\partial \mu_0} = -  \frac{3}{\pi}\int_{(m_1+m_2)^2}^\infty\text{d}s'\,\frac{\mu_0 \text{Im}\,\mathcal A(s',t)}{{(s'+\mu_0)}^4 } < 0 \, .\ee
Defining the IR subtraction constant at the low scale $\mu_0 = M_{\rm IR}^2$ and the UV at the high scale $\mu_0=M_{\rm UV}^2$ then we infer
\be
b_2^{\rm IR}(t) = b_2^{\rm UV}(t) +\frac{1}{\pi}\int_{(m_1+m_2)^2}^\infty\text{d}s'\, {\text{Im}\,\mathcal A(s',t)} \( \frac{1}{(s'+M_{\rm IR})^3 } - \frac{1}{(s'+M_{\rm UV})^3 } \) \, ,\ee
which ensures
\be
b_2^{\rm IR}(t) > b_2^{\rm UV}(t) \, ,
\ee
for small positive $t$. At a practical level this amounts to the fact that explicit contributions to the positivity bounds from physics at intermediate scales are necessarily positive so that the IR term is always larger than the UV term. Of course, for sufficiently negative coefficients in the UV, a positive spectral flow may still lead to a negative IR coefficient. So while the spectral flow is indicative, by itself this argument is not sufficient.

\subsection{Improved positivity bounds}\label{improved}

Returning to the standard case with only states with $J<2$, it is clear from the relation \eqref{eq:optical1} that the optical theorem carries much more information than what has been used to infer the positivity bound \eqref{positivity}. In particular, it is clear from the relation \eqref{eq:optical1} that the contribution from \emph{each} of the possible intermediate scattering processes gives a positive contribution to the imaginary part of the amplitude. We can make use of this fact by moving known contributions from the right-hand side of the theorem to the left-hand side, thus making the positivity bounds stronger and in fact generating further bounds on the couplings in the low-energy EFT. These go under the name of the improved positivity bounds \cite{deRham:2017xox,deRham:2017imi,Bellazzini:2016xrt} and will be relevant in the discussion below when comparing different contributions to the $2 - 2$ scalar scatterings.
Specifically we can separate the total imaginary part into two separately positive contributions
\be
\text{Im}\,\mathcal A(s,t)=\text{Im}\,\mathcal A_{\rm known}(s,t) + \text{Im}\,\mathcal A_{\rm unknown}(s,t)\,,
\ee
and further defining
\be
\tilde{\mathcal A}^{\rm imp}(s,t) \equiv \tilde{\mathcal A}(s,t) - \frac{s^2}{\pi}\int_{(m_1+m_2)^2}^\infty\text{d}s'\,\frac{\text{Im}\,\mathcal A_{\rm known}(s',t)}{{s'}^2 (s'-s)}-\frac{u^2}{\pi}\int_{(m_1+m_2)^2}^\infty\text{d}s'\,\frac{\text{Im}\,\mathcal A_{\rm known}(s',t)}{{s'}^2 (s'-u)}\,.
\ee
Then from \eqref{dispersion} the improved amplitude $\hat{\mathcal A}^{\rm imp}(s,t)$ satisfies its own dispersion relation
\be\label{improveddispersion}
\tilde{\mathcal A}^{\rm imp}(s,t) = a_0(t) + a_1(t) s + \frac{s^2}{\pi}\int_{(m_1+m_2)^2}^\infty\text{d}s'\,\frac{\text{Im}\,\mathcal A_{\rm unknown}(s',t)}{{s'}^2 (s'-s)}+\frac{u^2}{\pi}\int_{(m_1+m_2)^2}^\infty\text{d}s'\,\frac{\text{Im}\,\mathcal A_{\rm unknown}(s',t)}{{s'}^2 (s'-u)} \, ,
\ee
from which follows the first of many {\bf improved positivity bounds}
\be\label{improvedpositivity}
\frac{\text{d}^2\tilde{\mathcal A}^{\rm imp}(s,0) }{\text{d}s^2}>0\,,\,  \quad (m_1-m_2)^2  <s < (m_1+m_2)^2 \, .
\ee
For instance, one obvious application is to take as `known' the loop contributions calculated within the low-energy effective theory, valid for $s<\epsilon^2 \Lambda_c^2$ where $\Lambda_c$ is the cutoff of the effective theory, and $\epsilon \ll 1$ taken small enough that we can reliably trust the calculations, with `unknown' --- the remainder, i.e.
\be
\text{Im}\,\mathcal A_{\rm known}(s,t) =\theta(\epsilon^2 \Lambda_c^2 -s) \text{Im}\,\mathcal A(s,t) \, , \quad \text{Im}\,\mathcal A_{\rm unknown}(s,t) =\theta(s-\epsilon^2 \Lambda_c^2 ) \text{Im}\,\mathcal A(s,t) \, .
\ee
In the present context of a renormalizable theory coupled to gravity we have processes involving both gravitational and non-gravitational intermediate states. The improved positivity bounds, assuming applicability once the $t$-pole is removed, will enable us to remove the known non-gravitational contributions and focus instead on scatterings involving graviton exchange only. We refer the reader to \cite{newpaper} for a detailed analysis of the implications of the improved positivity bounds in the context of QED and the low-energy Euler--Heisenberg action.

\subsection{Compactified positivity bounds}

We have already highlighted the difficulties in applying the positivity bound \eqref{positivity} to amplitudes that manifest a massless $t$-channel pole with residue growing as $s^2$ (or faster), as is typically the case in processes that involve a graviton exchange.  In dealing with such a pole,  Ref.~\cite{Bellazzini:2019xts} provided a novel procedure for regularizing scattering amplitudes involving a massless graviton exchange. By compactifying one of the spatial directions on a circle, the initial 4d scattering process can effectively be reduced to a 3d one. Since in 3d the graviton is non-dynamical there can be no {\it physical} $t$-pole in the amplitude and thus naively no obstacle in applying the standard positivity bounds leading to constraints on the couplings in the original 4d low-energy action. In practise, a $t$-channel pole does remain perturbatively but as reviewed in section~\ref{Infrared} is expected to be removed on resummation, as is apparent in the eikonal approximation.
\\

In this work we shall check the consistency of these compactified positivity bounds by working with low-energy (IR) effective field theories for which the partial UV completion is known. We refer to appendix~\ref{sec:3d} for explicit details on the compactification process in our example. The advantage of this approach is the fact that the coefficients in the IR action are in fact determined by the partial UV theory and a direct comparison between these coefficients with the constraints that can be inferred from the compactified positivity bounds is possible. We will critically address issues with the compactification argument and positivity bounds applied in 3d in section~\ref{Infrared}.

\section{Scalar photon QED}\label{sec:qed}
In this work we shall deal with a simplified version of scalar QED where the photon is treated as a scalar field  \cite{Hollowood:2015elj}. An identical analysis can be also carried out for the vector Maxwell field, and spinor QED where the electron is a Dirac spinor \cite{newpaper}. Here we focus on a scalar field example since it  proves useful to highlight the apparent contradictions that arise when ignoring the $t$-channel pole due to the graviton exchange. We denote the ``scalar photon" by $\phi$, the ``scalar electron"  by $\psi$ and couple the two by a QED type of interaction $-\alpha M \phi\psi^2$, where $M$ is the mass of the heavy field $\psi$ and $\alpha$ is a dimensionless coupling constant. This scalar photon QED Lagrangian is
\be\label{scalar_photon}
\mathcal L_{\rm sQED}=\sqrt{-g}\left[-\frac{1}{2}(\partial\phi)^2-\frac{1}{2}(\partial\psi)^2-\frac{1}{2}M^2\psi^2-\alpha M \phi\psi^2\right]\,.
\ee
In distinction to the standard scalar QED with a massless vector field, ignoring its couplings to gravity, this is a super-renormalizable theory and thus has a different (and better) UV behaviour. Coupling this super-renormalizable field theory to gravity necessitates both the introduction of an Einstein--Hilbert term and covariant derivatives, together with non-renormalizable higher derivative interactions that require new UV physics at (or below) the Planck scale $\mpl$. We shall return to the impact of the latter corrections in section~\ref{sec:newUV}. For now we content ourselves with this theory minimally coupled to gravity
\be\label{scalar_photon2}
\mathcal L_{\rm pUV}=\sqrt{-g}\left[\frac{\mpl^2}{2}R-\frac{1}{2}(\partial\phi)^2-\frac{1}{2}(\partial\psi)^2-\frac{1}{2}M^2\psi^2-\alpha M \phi\psi^2\right]\,,
\ee
where pUV stands for partial UV completion.
It is straightforward to derive from \eqref{scalar_photon2} the low-energy effective theory that describes the dynamics of the photon well below the electron mass. Explicitly integrating out the massive electron in the presence of the gravitational field following the heat-kernel methods as in \cite{Drummond:1979pp,Shore:2002gw} (see \cite{Barvinsky:1987uw,Barvinsky:1990up,Barvinsky:1993en} for the original works) then leads to the low-energy effective action:
\be\label{toy}
\L_{\rm IR}^{(J)}=\sqrt{-g}\left[\frac{\mpl^2}{2}R-\frac{1}{2}(\partial\phi)^2-\frac{\alpha^3M}{(2\pi)^2}\frac{\phi^3}{3!}+\frac{\alpha^4}{2\pi^2}\frac{\phi^4}{4!}+C\frac{\alpha^2}{M^2}R^{\mu\nu}\partial_\mu\phi\,\partial_\nu\phi+\tilde C\frac{\alpha^4}{M^4}\(\partial\phi\)^4\right]\,,
\ee
with
\be\label{constC}
C = -\frac{1}{90(4\pi)^2}\,,\qquad\tilde C = \frac{1}{30(4\pi)^2}\,.
\ee
Let us stress that the sign of the coefficient $C$ in front of the $R^{\mu\nu}\partial_\mu\phi\partial_\nu\phi$ term is negative. This new interaction redresses the kinetic term of the scalar field and thus directly affects the propagation speed of the scalar $\phi$ on any gravitational background with a non-vanishing Ricci tensor. Intriguingly, the sign $C<0$ leads to superluminal low-energy propagation speeds relative to the background metric. It was shown in \cite{Hollowood:2015elj}, for a shockwave geometry, that despite the apparent causality violations in the low-energy theory, these are cleanly resolved in the high-frequency regime and the front velocity remains luminal. This ensures that the retarded propagator vanishes outside of the metric lightcone. The apparent low-energy superluminal phase and group velocity can never lead to any resolvable effect and is therefore never in tension with causality \cite{deRham:2020zyh}. \\

It is possible to rewrite the action \eqref{toy} in a more familiar form by performing a local field redefinition to remove the non-minimal couplings with the Ricci curvature. Naturally, such local transformations leave the scattering amplitudes invariant. Transforming the metric as
\be\label{trafo}
g_{\mu\nu}\to g_{\mu\nu}+2C\frac{\alpha^2}{M^2\mpl^2}\left(\partial_\mu\phi\,\partial_\nu\phi-\frac{1}{2}(\partial\phi)^2g_{\mu\nu}\right)\,,
\ee
leads to the Einstein-frame IR action
\be\label{IR_0}
\L_{\rm IR}=\sqrt{-g}\left[\frac{\mpl^2}{2}R-\frac{1}{2}(\partial\phi)^2-\frac{\alpha^3M}{(2\pi)^2}\frac{\phi^3}{3!}+\frac{\alpha^4}{2\pi^2}\frac{\phi^4}{4!}+\frac{\alpha^2}{M^2}\left(\frac{C}{\mpl^2}+\frac{\tilde C\alpha^2}{M^2}\right)\left(\partial\phi\right)^4+ \dots\right]\,,
\ee
up to subleading corrections.

\begin{figure}[t]
    \centering
\includegraphics[width=4cm]{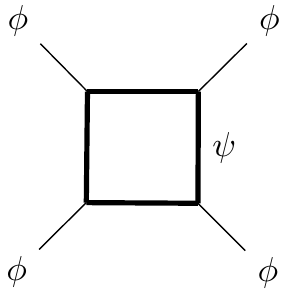}
\caption{Non--gravitational contribution to the $\phi\phi\to\phi\phi$ scattering in the UV theory. Bold lines represent propagators of the heavy field $\psi$.}
\label{fig:psibox}
\end{figure}

\paragraph{Compactified bounds:}
Consider now the tree-level $\phi\phi\to\phi\phi$ scattering as inferred from this low--energy EFT \eqref{IR_0}. The IR observer, only aware of the low-energy action \eqref{IR_0} with operator
\be
\mathcal L_{\rm IR}\supset \sqrt{-g}\frac{a}{M^4}\left(\partial\phi\right)^4\,,
\ee
with coupling $a$ would simply conclude that the coupling needs to be positive i.e. $a>0$. This specific statement has been long known for scalar field theories in the absence of gravity \cite{Pham:1985cr,Ananthanarayan:1994hf,Adams:2006sv}. It was presented again in \cite{Bellazzini:2019xts} where it was argued that $a>0$ should hold also in the presence of gravity for any weakly coupled\footnote{In this context we may take weakly coupled to mean calculable within a standard perturbative expansion.} UV completion of \eqref{IR_0} by means of the compactified positivity bounds. In the present context, the compactified positivity bounds then imply
\be\label{bound1}
\frac{C}{\mpl^2}+\frac{\tilde C\alpha^2}{M^2}>0\,.
\ee
Comparing to \eqref{constC} and naively using the fact that $M\ll \mpl$ we could conclude that this bound is always satisfied. Another way of seeing the bound above is to rewrite it as bound on the charge-to-mass ratio, $\alpha/M$, giving
\be\label{bound2}
\frac{\alpha}{M}>\frac{1}{\sqrt{3}}\frac{1}{\mpl}\,,
\ee
akin very much to similar arguments in QED made to support the weak gravity conjecture \cite{ArkaniHamed:2006dz,Cheung:2014ega}.
Thus it appears to be the case that the positivity bounds are satisfied provided that \eqref{bound2} is respected.
\\

This conclusion is however premature. The reason for this is the fact that the operator $\tilde C\frac{\alpha^4}{M^4}\left(\partial\phi\right)^4$ in the IR action \eqref{IR_0} arises from a non-gravitational loop diagram in the UV theory, shown in Fig.~\ref{fig:psibox} and computed in appendix~\ref{app:box}. Rather than working with the low-energy theory \eqref{IR_0} we may return to the partial UV completion \eqref{scalar_photon2} and apply the improved positivity bounds \cite{deRham:2017xox,deRham:2017imi,Bellazzini:2016xrt} reviewed in section~\ref{improved} to remove the contribution from the box diagram. From the low-energy point of view this amounts to removing the $\tilde C$ term, leading to a bound on the gravitational correction alone. This is considered in more detail in \cite{newpaper}. In order to avoid these technical subtleties, in what follows we shall focus on related amplitudes that violate compactified positivity without invoking the improved positivity bounds.


\section{Scalar photon QED with a spectator field}\label{sec:uv}

The discussion of the previous section can be considerably sharpened by including an additional light spectator field $\chi$  in the UV Lagrangian \eqref{scalar_photon2}. We assume that the spectator field $\chi$ interacts with the scalar QED theory only via gravity\footnote{The trick of introducing a spectator field to probe positivity bounds was heavily used in \cite{deRham:2019ctd}. As highlighted in section~\ref{sec:qed}, we can reach the same conclusions without introducing a spectator field, however this relies on implementing the improved positivity bounds. The inclusion of a spectator field is only performed so as to bypass this need of going through improved bounds. Indeed, when considering the elastic  $\chi\phi\to\chi\phi$ scattering there is no need to subtract other known contributions.}.   For simplicity but without loss of generality, we keep both of the scalar fields $\phi$ and $\chi$ massless here\footnote{One can in principle introduce a small mass $m$ for either of them for as long as $m^2/M^2\ll 1$. We shall do so in appendix~\ref{sec:addUV} when considering possible extensions to the action~\eqref{pUV_0} leading to processes with light scalar loops. However, having a non-zero scalar field mass does not provide any additional insights in the scattering process considered in the current section.}.
The scalar photon QED action minimally coupled to gravity and including a light spectator field $\chi$ takes the form:
\be\label{pUV_0}
\L_{\rm pUV}=\sqrt{-g}\left[\frac{\mpl^2}{2}R-\frac{1}{2}(\partial\chi)^2-\frac{1}{2}(\partial\phi)^2-\frac{1}{2}(\partial\psi)^2-\frac{1}{2}M^2\psi^2-\alpha M\phi\psi^2\right]\,.
\ee
We have assumed that the spectator $\chi$ is a scalar field here but it could in principle be any other (fermion or vector) field coupled to gravity. In this setup we can then focus on the elastic $\phi\chi\to\phi\chi$ scattering that is always mediated by the massless graviton exchange and only occurs via $t$-channel scattering. We shall provide more details in exploring the validity of the compactified positivity bounds and analyzing the $2-2$ scattering between the spectator and matter both in the partial UV and low-energy EFT as well as on the compactified manifold in the appendices.

\subsection{Scattering in the low-energy EFT}\label{sec:ir}

To begin with, we determine the low-energy effective theory associated with \eqref{pUV_0} that arises below the electron mass. This is straightforward following the approach of the previous section. Integrating out the scalar electron leads to the IR action
\be\label{toy2}
\L_{\rm IR}^{(J)}=\sqrt{-g}\left[\frac{\mpl^2}{2}R-\frac{1}{2}(\partial\chi)^2-\frac{1}{2}(\partial\phi)^2-\frac{\alpha^3M}{(2\pi)^2}\frac{\phi^3}{3!}+\frac{\alpha^4}{2\pi^2}\frac{\phi^4}{4!}+C\frac{\alpha^2}{M^2}R^{\mu\nu}\partial_\mu\phi\,\partial_\nu\phi+\tilde C\frac{\alpha^4}{M^4}\(\partial\phi\)^4\right]\,,
\ee
which on applying the field redefinition \eqref{trafo} is seen to be equivalent to
\ba\label{IR_1}
\L_{\rm IR}=\sqrt{-g}\Bigg[\frac{\mpl^2}{2}R-\frac{1}{2}(\partial\chi)^2-\frac{1}{2}(\partial\phi)^2-\frac{\alpha^3M}{(2\pi)^2}\frac{\phi^3}{3!}+\frac{\alpha^4}{2\pi^2}\frac{\phi^4}{4!}\ \\
+C'\frac{\alpha^2}{M^2\mpl^2}\left(\partial\phi\right)^4+C\frac{\alpha^2}{M^2\mpl^2}(\partial_\mu\phi\,\partial^\mu\chi)^2+ \dots \Bigg]\,, \nn
\ea
where we have neglected subleading terms and the coefficient $C'$ is given by
\be
\frac{C'\alpha^2}{M^2\mpl^2}\equiv\frac{\alpha^2}{M^2}\left(\frac{C}{\mpl^2}+\frac{\tilde C\alpha^2}{M^2}\right)\,.
\ee
The new feature is the contact interaction $(\partial_\mu\phi\,\partial^\mu\chi)^2$ which contributes directly to the $\chi\phi\to\chi\phi$ scattering. As usual the scattering amplitude is a physical quantity and is independent on the field redefinitions that we have performed to get to the form of the IR action as in \eqref{IR_1}. The field redefinition is merely used as a tool to make this interaction more explicit. \\

 \begin{figure}[t]
    \centering
\includegraphics[height=1.3in]{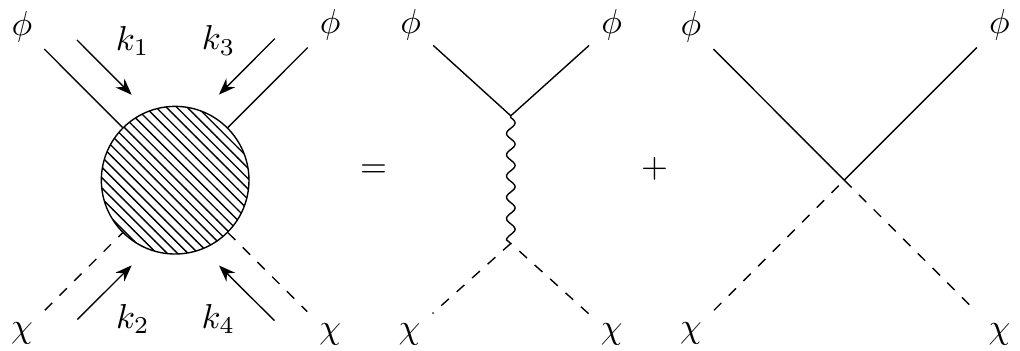}
\caption{Contributions to $\chi\phi\to\chi\phi$ scattering in the IR theory. The wiggly line represents the graviton propagator.}
\label{fig:scattering2}
\end{figure}

At the leading order the $\chi\phi\to\chi\phi$ scattering in the IR theory occurs via two different scattering processes shown in Fig.~\ref{fig:scattering2}. As compared to the previous case, there is a new contact interaction directly between the two fields $\phi$ and $\chi$. For the leading order scattering amplitude we then find
\be\label{amp_IR}
\mathcal A_{\rm IR}(s,t)= -\frac{s(s+t)}{\mpl^2t} +\frac{C\alpha^2}{M^2\mpl^2}\left[s(s+t)+\frac{t^2}{2}\right]\,.
\ee
As argued in \cite{Bellazzini:2019xts} and as shown in appendix~\ref{sec:3d} applying the compactified forward limit positivity bounds for this amplitude amounts to dropping the $t$-pole and requiring that
\be
\frac{\text{d}^2\mathcal A_{\rm IR, no\,pole}(s,0)}{\text{d}s^2}=\frac{2C\alpha^2}{M^2\mpl^2}>0\,,
\ee
leading to the requirement that $C>0$. In other words, the new positivity bounds state that for this IR theory \eqref{IR_1} to have a standard (local and Lorentz invariant) UV completion the coupling constant $C$ has to be positive.
However, comparing \eqref{amp_IR} to the corresponding amplitude \eqref{amp_pUV} computed below in the UV theory we read off
\be\label{constC2}
C=-\frac{1}{90(4\pi)^2}<0\,,
\ee
which contradicts the positivity bound above. It shows that a theory can have an apparently healthy partial UV completion (given in \eqref{pUV_0}) even though the compactified positivity bounds are violated.

\subsection{Scattering in the UV completion}

In the previous subsection we have seen that on introducing a spectator field into the partial UV completion, the compactified positivity bounds applied to the low-energy effective theory are violated, and no condition analogous to \eqref{bound2} can be imposed to save them. Thus \eqref{pUV_0} is a particularly straightforward example of a partial UV completion that would be expected to be well-defined, but for which compactified positivity bounds are violated. In section \ref{sec:newUV} we will discuss possible corrections to the partial UV  completion that may rectify this. For now we would like to see how positivity bounds applied directly to the UV theory  \eqref{pUV_0}  play out.\\

\begin{figure}[t]
    \centering
\includegraphics[width=12cm]{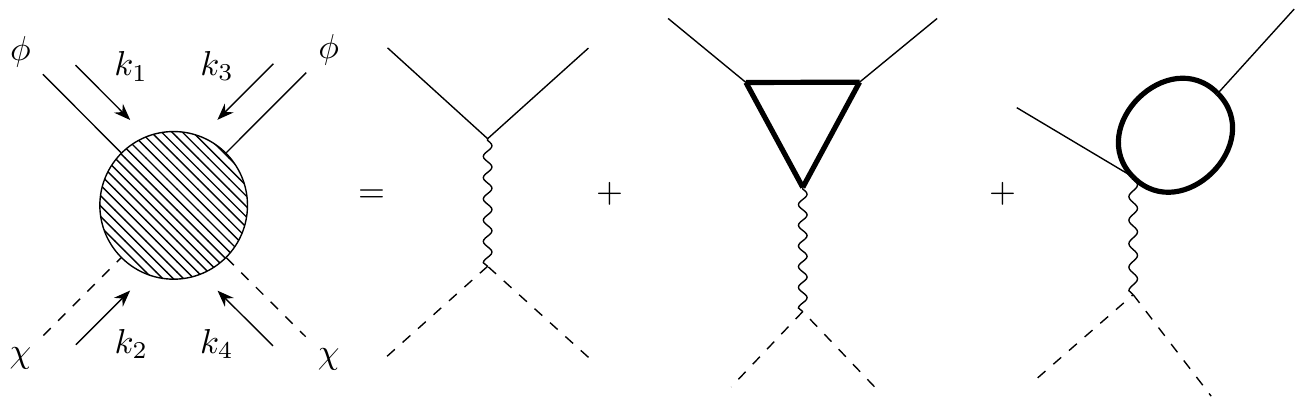}
\caption{Leading contributions to the $\chi\phi\to\chi\phi$ scattering as seen from the point of view of the partial UV theory. Wiggly lines stand for the graviton propagator while thick lines stand for the propagator of the heavy field $\psi$. }
\label{fig:scattering}
\end{figure}

This calculation may be performed in two different ways. We may first compute the scattering amplitude in the partial UV completion in 4d Minkowski $\mathbb{R}^{3,1}$, and use this to match against the low-energy effective theory to confirm the results in the previous section. The compactified positivity bounds may then be applied directly in the IR EFT as per the previous section. Alternatively we may apply the compactification procedure directly to the partial UV completion \eqref{pUV_0}. These two procedures are not identical because in the former, the electron loop is computed in 4d Minkowski $\mathbb{R}^{3,1}$, whereas in the latter it is computed on a compactified space $\mathbb{R}^{2,1} \times S^1$. Performing the calculation both ways allows us to test whether there is any issue in running the argument due to failure of the limits $L \rightarrow \infty$ and the low-energy limit $E/M \rightarrow 0$ to commute. The former calculation implicitly assumes $M_{\rm KK} \ll M$, whereas the latter is applicable also when $M_{\rm KK} \gg M$. We will save the technical details for appendix~\ref{sec:uv_ampl} for the $\mathbb{R}^{3,1}$ calculation and appendix~\ref{sec:3d} for the $\mathbb{R}^{2,1} \times S^1$ and review the essential results here.

\subsubsection{Electron loops in 4d Minkowski $\mathbb{R}^{3,1}$}

The contributions to the tree-level and one-loop $\chi\phi\to\chi\phi$ scattering amplitude $\mathcal A_{\rm pUV}(s,t)$ relevant for the positivity bounds in the partially UV complete theory \eqref{pUV_0} come from the diagrams given in Fig.~\ref{fig:scattering}, and in the notations of appendix~\ref{sec:uv_ampl} are
\be
\mathcal A_{\rm pUV}(s,t)=Z_\phi\times\mathcal A_{\rm tree,0}+\mathcal A_{\phi\psi^2}+\mathcal A_{\phi\psi^2h} \, .
\ee
Given the wave function renormalization
\be\label{wavefunction}
Z_\phi=1-\frac{\alpha^2}{3 (4 \pi)^2}\,,
\ee
 explicitly this is
\ba\label{amp_pUV10}
\mathcal A_{\rm pUV}(s,t) =  \frac{s u}{\mpl^2t} +\Bigg[
&-&\frac{\alpha^2}{3 (4 \pi)^2} \frac{s u}{\mpl^2t}\\
&+&\frac{8M^2\alpha^2}{\mpl^2t}\int_0^1\text{d}x\int_0^{1-x}\text{d}y\left\{\frac{A(s,t)}{32\pi^2\Delta}+\frac{t}{32\pi^2}\left(-1-\log\frac{\Delta}{M^2}\right)\right\}\Bigg]
\,,\nn
\ea
with $A(s,t) \equiv -s(s+t)(-1+x+y)^2+M^2t$ and $\Delta\equiv M^2-xyt$.
Let us note that, as expected, this amplitude is independent of the renormalization scale. To see this cancellation
explicitly we have followed the same procedure as highlighted in appendix~\ref{sec:phi3}.
The role of the wavefunction renormalization is to ensure that the terms in square brackets contain no net contribution to the $t$-pole. This confirms the equivalence principle at the quantum level, namely that the graviton only couples to the physical energy at the strength $\mpl$.
The remaining part of the amplitude contains no $s$-channel branch cut or pole, i.e. there is no imaginary part for physical values of $s$. This is the virtue of introducing the spectator field whose interactions only arise through gravity. Hence there are no dispersive imaginary parts to remove via improved positivity bounds and we may just apply the usual positivity bounds.

Expanding at low-energy, we confirm the amplitude determined from the low-energy effective theory as expected
\be\label{amp_pUV}
\mathcal A_{\rm pUV}(s,t)=-\frac{s^2}{\mpl^2t}-\frac{\alpha^2s^2}{90(4\pi)^2M^2\mpl^2}+\mathcal O(t^0)\,,
\ee
where $\mathcal O(t^0)$ stands for all other contributions to the amplitude that are finite, non-singular and have no $s^2$ dependence. Applying the prescription suggested in Ref.~\cite{Bellazzini:2019xts} allows us to ignore the $t$-pole so that the positivity bound of Eq.~\eqref{positivity} implies
\be\label{contradiction}
\frac{\text{d}^2 \mathcal A_{\rm pUV, no\,pole}(s,0)}{\text{d}s^2}=-\frac{2\alpha^2}{90(4\pi)^2M^2\mpl^2}>0\,,
\ee
which cannot be satisfied for any real values of the coupling $\alpha$.\\

Closer inspection of \eqref{amp_pUV10} shows that even with the $t$-channel pole removed, the remaining part of the amplitude in square brackets grows as $s^2$ for fixed $t$ at large $s$, thus invalidating the assumption $\lim_{|s| \rightarrow \infty} s^{-2} \A(s,t) \rightarrow 0$. This should not come as a surprise since  \eqref{pUV_0} in itself only a partial UV completion by virtue of being gravitational, and new physics will come in at higher energies to resolve this. How precisely this new physics manifests itself i.e. whether new physics enters through higher spins or through other fields is irrelevant to this point,  the central issue is rather about the scale at which  this new physics enters. The naive expectation would be that the cutoff is the Planck scale, or the string scale, but as we shall see the cutoff ought to be much lower for the compactified positivity bounds to be satisfied.  This possibility is indeed explored in section~\ref{sec:newUV} where we show that in order for our partial UV completion \eqref{pUV_0} to admit a full UV completion for which the compactified positivity bounds apply, new physics has to be introduced at a scale $\Lambda$ significantly lower than the Planck scale $\Lambda \sim (M \mpl)^{1/2}$. From the EFT perspective, we know that this result is obtained in a naively sensible effective field theory \eqref{pUV_0} where the EFT cutoff --- induced by the presence of the Einstein--Hilbert term and the minimal coupling to gravity --- would naively be expected to coincide with the Planck scale, or a scale close to it. Given the simplicity of the model in \eqref{pUV_0} and its resemblance to the standard QED this would be a very strong statement with far-reaching implications to even Standard Model physics. It seems more likely that these findings point to some inconsistency in the handling of the $t$-channel pole.

\subsubsection{Electron loops in the compactified $\mathbb{R}^{2,1} \times S^1$ space}

In order to make sure that the simplified prescription of dropping the $t$-pole does indeed follow from the regularization procedure suggested in~\cite{Bellazzini:2019xts} we shall repeat the same strategy to the partial UV completion \eqref{pUV_0}. The technical details are found in appendix~\ref{sec:3d}. This calculation is most straightforwardly viewed by compactifying the 4d action while maintaining the contribution from the Kaluza-Klein (KK) modes that contribute to the loop processes. To the order needed, the effective 3d action obtained from compactification on a $S^1$ of length $L$ is
\ba\label{3Deff}
S_{\rm UV, 3d}&\supset L&\int \d^3x\sqrt{-g}\Bigg\{\frac{\mpl^2}{2}R-\frac{1}{2}(\partial\chi_0)^2-\frac{1}{2}(\partial\phi_0)^2-\frac{1}{2}(\partial\psi_0)^2-\frac{1}{2}M^2\psi_0^2\\
&-&\alpha M \phi_0\psi_0^2-2\alpha M\sum_{n=1}^\infty  \phi_0\psi_n\psi_{n}^\dagger-\sum_{n=1}^\infty \left(g^{\mu\nu}\partial_\mu\psi_n\partial_\nu\psi_{n}^\dagger+\left(M^2+\frac{4\pi^2n^2}{L^2}\right)\psi_n\psi^\dagger_n\right)\Bigg\}\,,\nn
\ea
where $\psi_0$, $\phi_0$ and $\chi_0$ are the KK zero modes and $\psi_n$ the KK modes with masses $M_n^2=M^2+\frac{4\pi^2n^2}{L^2}$. The zero mode elastic scattering amplitude $\chi_0 \phi_0 \rightarrow \chi_0 \phi_0 $ is determined to one-loop order by the analogous Feynman diagrams in Figs.~\ref{fig:scattering3} and \ref{fig:KKdiagram} in appendix~\ref{sec:3d}.
The structure of the one-loop amplitude is similar
\be
\mathcal A_{\rm pUV,3d}(s,t)=Z_{\phi_0} \times\mathcal A_{\rm tree,0}+\sum_{n=0}^{\infty} \mathcal A_{\phi_0\psi_n^2}+\sum_{n=0}^{\infty} \mathcal A_{\phi_0\psi_n^2h_0} \, .
\ee
The wavefunction renormalization factor becomes
\be
Z_{\phi_0}=1-\frac{\alpha^2M^2}{48\pi L}\sum_{n=-\infty}^\infty\frac{1}{M_n^3}\,,
\ee
which is equivalent to \eqref{wavefunction} in the limit $L \rightarrow \infty$. As before this ensures the equivalence principle at the quantum level, and the low-energy expansion of the 3d amplitude is
\be
\mathcal A_{\rm pUV, 3d}(s,t)=-\frac{s^2}{\mpl^2Lt}-\frac{\alpha^2M^2}{240(4\pi)L^2\mpl^2}\sum_{n=-\infty}^\infty \frac{s^2}{M_n^5}+\mathcal O(t^0)\,.
\ee
Assuming the validity of the eikonal argument that allows us to remove the $t$-channel pole as argued in~\cite{Bellazzini:2019xts}, we would be led to the compactified positivity bound
\be
\frac{\text{d}^2\mathcal A_{\rm pUV, 3d}(s,0)}{\text{d}s^2}=-\frac{\hat \alpha^2}{120(4\pi)M M_3}\times \frac{1}{M^2}\left(1+2\sum_{n=1}^\infty\frac{1}{\left(1+\frac{4\pi^2n^2}{L^2M^2}\right)^{5/2}}\right)>0\,,
\ee
where we have defined the 3d Planck mass $M_3= \mpl^2L$ and effective 3d coupling $\hat \alpha = \alpha/\sqrt{L}$. It is now clear that this bound is violated for any $L$. In the limit $L\rightarrow \infty$ it reproduces \eqref{contradiction}
\be
\lim_{L \rightarrow \infty} L \frac{\text{d}^2\mathcal A_{\rm pUV, 3d}(s,0)}{\text{d}s^2}=-\frac{2\alpha^2}{90(4\pi)^2M^2\mpl^2}>0\,,
\ee
whereas in the limit $L \rightarrow 0$, only the zero mode contributes in the loop and we obtain the pure 3d result
\be
\lim_{L \rightarrow 0} \frac{\text{d}^2\mathcal A_{\rm pUV, 3d}(s,0)}{\text{d}s^2}=-\frac{\hat \alpha^2}{120(4\pi)M^3 M_3}>0\,,
\ee
that would result from the pure 3d action
\ba\label{3Deff2}
S'_{\rm UV, 3d}= \int \d^3x\sqrt{-g}\Bigg\{\frac{M_3}{2}R-\frac{1}{2}(\partial \hat \chi)^2-\frac{1}{2}(\partial \hat \phi)^2-\frac{1}{2}(\partial \hat \psi)^2-\frac{1}{2}M^2\hat \psi^2
-\hat \alpha M \hat \phi \hat \psi^2\Bigg\}\,,
\ea
which is obtained from \eqref{3Deff} by canonically normalizing $\psi_0 = \hat \psi/\sqrt{L}$, $\chi_0 = \hat \chi/\sqrt{L}$, $\phi_0 = \hat \phi/\sqrt{L}$ and sending $L \rightarrow 0$, for fixed $\hat \alpha$, to decouple the KK modes. The partial UV completion \eqref{3Deff2} is itself super-renormalizable in the absence of gravity and by itself serves as an example of a partial UV completion where 3d positivity bounds are violated.

\section{New UV physics}\label{sec:newUV}

The physical motivation of the specific interactions in our UV theory \eqref{pUV_0} is clearly that of its resemblance to the standard QED. As our computations in section \ref{sec:uv} clearly show, if we were to apply the compactified positivity bounds of Ref.~\cite{Bellazzini:2019xts} to the IR theory \eqref{IR_1} we would conclude that the theory only admits a standard UV completion if the coefficient $C$ is positive. On the other hand the known partial UV completion \eqref{pUV_0} gives $C<0$ in Eq.~\eqref{constC2}. As already alluded to, there may be different reasons for this disagreement:
\begin{enumerate}[label=(\alph*)]
\item Either the compactified positivity bounds do not generically apply. This could be due to many reasons: One issue highlighted in \cite{Loges:2019jzs} is that the argument relies on taking the forward limit before the large $L$ limit. However as we have seen in the previous section, positivity is violated for any $L$ for $t \sim 0$ and the 3d theory \eqref{3Deff2} by itself violates positivity. This suggests that any problems relate directly to the subtleties intrinsic to 3d gravitational theories and 3d scattering amplitudes. We will highlight some of these in the next section \ref{Infrared}.

\item Or the compactified positivity bounds are indeed always valid and in this context they imply that by itself the partial UV completion \eqref{pUV_0} (or \eqref{scalar_photon2}) minimally coupled to gravity is not a valid one, at least not without the existence of new UV physics at a scale parametrically lower than the Planck scale.

\end{enumerate}
In this section we shall consider the second possibility. There are two relatively straightforward possibilities for enhancing the partial UV theory in attempting to satisfy the new positivity bounds. Both rely on introducing interactions already at the level of the partial UV theory, thus providing a different UV completion from the one presented in \eqref{pUV_0}. The distinction between the two possibilities is whether the new interactions would be renormalizable or non-renormalizable in the absence of gravity.

\subsection{Renormalizable operators}

In the actual QED case there are of course no further renormalizable interactions in 4 dimensions. However, in our scalar photon QED model \eqref{pUV_0} we could supplement \eqref{pUV_0} with the additional terms
\be
\Delta {\cal L} = \sqrt{-g} \left[ -aM\frac{\phi^3}{3!} - bM\psi\frac{\phi^2}{2}-\frac{\lambda}{4!}\phi^4 - \frac{\gamma}{4}\psi^2\phi^2+d_1 \psi^3 + d_2 \phi \psi^3 + d_3 \psi^4 + d_4 \phi^3 \psi \right]\,,
\ee
each of which is renormalizable in the absence of gravity. The terms with coefficients $d_i$ will not contribute to $\chi\phi\to\chi\phi$ scattering at one-loop order and we do not consider them any further in what follows. The remaining interactions are discussed in great detail in appendix~\ref{sec:addUV}. As before in all cases once the wavefunction renormalization is accounted for, the $t$-pole is universally $-s^2/(\mpl^2 t)$. The additional contributions to the positivity bound are
\be
 \frac{\text{d}^2 \Delta\mathcal A_{\rm pUV, 3d}(0,0)}{\text{d}s^2} = -\frac{a^2M^2}{16\pi^2}\frac{1}{\mpl^2}\frac{(45-8\sqrt{3} \pi)}{162m_{\rm ph}^4}-\frac{b^2}{9 (4\pi)^2M^2\mpl^2}\,.
\ee
For any choice of coefficients $a,b$ these contributions are negative and so cannot solve the contradiction. Thus we see that quite generically renormalizable field theories minimally coupled to gravity tend to violate the compactified positivity bounds. \\

We could also imagine introducing additional renormalizable interactions between the spectator $\chi$ and the other matter fields $\phi$ and $\psi$. These are considered in detail in appendix~\ref{app:addmore}. The story is quite analogous, all contributions to the positivity bounds are negative except that from the box diagram pictured in Fig.~\ref{fig:uv2} in appendix~\ref{app:addmore}. However the reason this gives a positive contribution is that it is a non-gravitational contribution that itself satisfies the Froissart bound. Hence it may be removed by application of the improved positivity bounds. We thus find that no introduction of renormalizable interactions, even with the spectator field, resolves the contradiction.

\subsection{Non-renormalizable operators}\label{sec:nonren}

We are thus led to looking at modifying the partial UV completion by introducing new physics at a higher energy scale $\Lambda \gg M$. For instance this may be new physics at the Planck scale, or a lower scale such as the string scale. In the latter case the UV completion may require the introduction of a tower of higher spin Regge states \cite{Hamada:2018dde}. Regardless of what this new physics is, it will show up at low energies as irrelevant operators extending the partial UV completion \eqref{pUV_0}. Indeed, it is straightforward to introduce new irrelevant operators so as to satisfy the compactified positivity bounds. The price to pay is the fact that these irrelevant interactions must come in at a scale much lower than the Planck scale --- hence significantly reducing the cutoff of the partial UV theory \eqref{pUV_0}. We discuss this case in detail in this section.

The most obvious way to ensure the compactified positivity bounds are satisfied is to supplement that partial UV completion  \eqref{pUV_0} with the irrelevant operator
\be\label{corr1}
\Delta {\cal L} = \sqrt{-g} \, B R^{\mu\nu} \p_{\mu} \phi \p_{\nu} \phi \, ,
\ee
where $B$ is a dimensionful coupling. On integrating out the $\psi$ loop, the net contribution to the IR action \eqref{toy2} from this interaction will then by
\be
\L_{\rm IR}^{(J)} = \sqrt{-g} \(B+ C \frac{\alpha^2}{M^2} \) R^{\mu\nu} \p_{\mu} \phi \p_{\nu} \phi \, ,
\ee
and so by choosing
\be
B > - C \frac{\alpha^2}{M^2}\,,
\ee
the compactified positivity bound will be satisfied. The problem is since this is an irrelevant operator it will induce a new cutoff in the effective theory. To determine this cutoff it is helpful to perform in the partial UV completion \eqref{pUV_0} with the additional operator \eqref{corr1} a field redefinition analogous to~\eqref{trafo}
\be\label{trafo2}
g_{\mu\nu}\to g_{\mu\nu}+\frac{2B}{\mpl^2}\left(\partial_\mu\phi\,\partial_\nu\phi-\frac{1}{2}(\partial\phi)^2g_{\mu\nu}\right)\,,
\ee
which generates corrections to \eqref{pUV_0} of the form
\be
\Delta {\cal L}' = \frac{B}{\mpl^2} \( (\partial_\mu\phi\,\partial^\mu\phi)^2+ (\partial_\mu\phi\,\partial^\mu\psi)^2+(\partial_\mu\phi\,\partial^\mu\chi)^2 \) + \dots\,.
\ee
These obviously give another scattering channel contributing to the $\chi\phi\to\chi\phi$ scattering amplitude in the UV theory allowing to satisfy the positivity bound as stated above.
Assuming $\alpha \sim {\cal O}(1)$ then this is an irrelevant operator\footnote{A priori the irrelevant operator contains a term behaving as $\p^2 h (\p \phi)^2/(\mpl M^2)$ which would suggest an even lower cutoff however this particular term is removable by a field redefinition and does not enter the amplitude. Ultimately the irrelevant operator that has to be included in the EFT ought to affect the amplitude at the scale $\Lambda \sim (M \mpl)^{1/2}$.} of the form $\mpl^2 R (\p \phi)^2/\Lambda^4$ with cutoff of at most $\Lambda \sim (M \mpl)^{1/2}$.

Stated differently, we could add to \eqref{pUV_0} generic non-renormalizable operators appearing at some scale $\Lambda$:
\be\label{pUV_3}
\begin{split}
\L_{\rm pUV,2}&=\sqrt{-g}\Big[\frac{\mpl^2}{2}R-\frac{1}{2}(\partial\chi)^2-\frac{1}{2}(\partial\phi)^2-\frac{1}{2}(\partial\psi)^2-\frac{1}{2}M^2\psi^2\\
&\qquad-\alpha M\phi\psi^2+\frac{\gamma}{\Lambda^4}(\partial_\mu\chi\partial^\mu\psi)^2+\frac{\delta}{\Lambda^4}(\partial_\mu\chi\partial^\mu\phi)^2
+\frac{\beta}{\Lambda^4}(\p  \phi)^4+\frac{\sigma}{\Lambda^4}(\partial_\mu\phi\partial^\mu\psi)^2+ \dots
\Big]\,.
\end{split}
\ee
\begin{figure}
\centering
\begin{subfigure}{0.35\textwidth}
\centering
\includegraphics[height=1.2in]{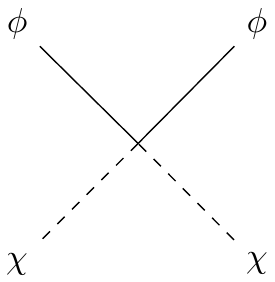}
\caption{Tree-level contribution from $\mathcal A_{\phi^2\chi^2}$} \label{fig:uv3}
\end{subfigure}
\hspace*{0.5in} 
\begin{subfigure}{0.35\textwidth}
\centering
\includegraphics[height=1.2in]{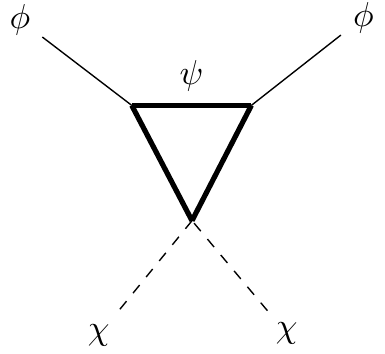}
\caption{One-loop contribution from $\mathcal A_{\psi^2\chi^2}$} \label{fig:uv4}
\end{subfigure}
\caption{Feynman diagrams contributing to $\chi\phi\to\chi\phi$ scattering from the new UV operators in \eqref{pUV_3} up to one loop. Bold lines correspond to the propagators of the heavy field $\psi$.} \label{fig:newUV}
\end{figure}
Terms such as $(\partial\chi)^2(\partial\psi)^2$ will not contribute to $s$ dependence of $\chi\phi\to\chi\phi$ scattering and may be ignored. The new diagrams relevant for the one-loop $\chi\phi\to\chi\phi$ scattering are shown in Fig.~\ref{fig:newUV}:
\begin{itemize}
\item The diagram in Fig.~\ref{fig:uv3} is a simple contact diagram that we have already computed in section \ref{sec:ir} with the corresponding vertex $V_{\phi^2\chi^2}$ given in \eqref{rules}.  For the scattering amplitude in the forward limit it gives
\be
\mathcal A_{\phi^2\chi^2}(s,0)= \frac{\delta}{\Lambda^4}s^2\,.
\ee
Hence, its contribution is positive for $\delta>0$. As discussed it can be used to cancel the negative $s^2$ contribution to the $\chi\phi\to\chi\phi$ scattering amplitude provided
\be
\frac{\delta}{\Lambda^4}>\frac{\alpha^2}{90(4\pi)^2M^2\mpl^2}\,.
\ee
Importantly, this implies that the scale of the new non-renormalizable interaction that we are adding in \eqref{pUV_3} has to be $\Lambda\leq(M\mpl)^{1/2}\ll \mpl$. Hence, the new positivity bounds can be satisfied in a UV theory with new derivative interactions that become important at a scale $\Lambda\ll\mpl$. Notably, in comparison to our initial UV theory \eqref{pUV_0} this means that the EFT cutoff decreases from $\mpl$ to $\Lambda$. We will discuss the implications of this low cutoff in more detail below.

\item The diagram in Fig.~\ref{fig:uv4} is new. The corresponding interaction in \eqref{pUV_3} gives the same vertex, $V_{\psi^2\chi^2}$ as in the previous step with $\phi\leftrightarrow\psi$ and $\delta\leftrightarrow\gamma$.
  The amplitude of the process shown is
\be
i\mathcal A_{\psi^2\chi^2}(s,t)=\int\frac{\d^dp}{(2\pi)^d}V_{\phi\psi^2}^2V_{\psi^2\chi^2}\Delta_\psi(p)\Delta_\psi(k_3+p)\Delta_\psi(p-k_1)\,.
\ee
It has no $t$-channel pole and one can evaluate its $s^2$ contribution in forward limit by directly setting $t=0$. We find
\be
i\mathcal A_{\psi^2\chi^2}(s,0)=\frac{16\alpha^2\gamma M^2}{\Lambda^4}\int \frac{\d^dk}{(2\pi)^d}\int_0^1\text{d}x\int_0^{1-x}\text{d}y\,\frac{\frac{1}{2}s^2(-1+x+y)^2}{[k^2+M^2]^3}\,.
\ee
After integrating over momentum we are left with an integral that we have already performed in \eqref{tpole1}. For the amplitude we thus get
\be
\mathcal A_{\psi^2\chi^2}(s,0)=\frac{\alpha^2\gamma}{48\pi^2\Lambda^4}\times s^2\,.
\ee
Comparing this with the scattering amplitude computed in the initial UV theory we find again that in order for the total contribution to the $s^2$ term to be positive the EFT cutoff scale should be $\Lambda\leq(M\mpl)^{1/2}\ll \mpl$.
\item The operators considered in \eqref{pUV_3} contribute to the $\chi \phi \to \chi \phi$ through other diagrams at up to one--loop but all other contributions are suppressed compared with the ones mentioned previously and depicted in Fig.~\ref{fig:newUV}. Relying on those to ensure that the compactified positivity bounds are satisfied for the $\chi \phi \to \chi \phi$ amplitude would only lead to an even smaller cutoff.
\end{itemize}

\begin{figure}[t]
    \centering
\includegraphics[width=3cm]{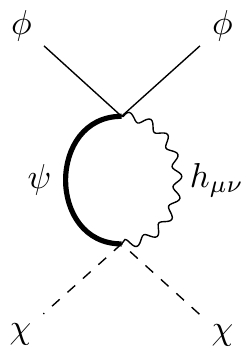}
\caption{One-loop contribution to $\chi\phi\to\chi\phi$ scattering from cubic dimension 5 operators. The wiggly line represents the graviton propagator, while the bold line that of the heavy field $\psi$.}
\label{fig:cubicUV}
\end{figure}

We should also account for the role of irrelevant cubic operators. For instance the dimension 5 operators
\be
\Delta \L_{\rm pUV,2} = \sqrt{-g} \left(  \frac{f_1}{\Lambda} ( \partial \phi)^2 \psi +   \frac{f_2}{\Lambda} ( \partial \chi)^2 \psi  \right)\,,
\ee
will contribute to the amplitude in two ways, one through the process in Fig.~\ref{fig:cubicUV} mediated partly by graviton exchange within a loop. On dimensional grounds this contributes to the positivity bounds by an amount
\be
\Delta \frac{\text{d}^2}{\text{d}s^2}\mathcal A_{\rm pUV}(0,0) \sim \frac{f_1f_2}{\mpl^2 \Lambda^2} \, .
\ee
While the sign can be engineered to be positive, for this to compensate the negative contributions, we would need the cutoff at the scale $\Lambda \sim M$ which is worse than what is needed for quartic interactions. In addition this will lead to a simple tree-level $\psi$ exchange process but this does not give rise to any $s$ dependence.

\subsection{$\phi\phi\to\phi\phi$ scattering}

So far we have only explored the compactified positivity bounds for the  $\chi \phi \to \chi \phi$ amplitude as the presence of the spectator considerably simplifies the discussion. However as discussed in section~\ref{sec:qed}, applying a similar type of 3d regularization argument to the improved positivity bounds for the $\phi \phi \to \phi \phi$ scattering would lead to a similar contradiction. Attempting to resolve this effect through the introduction of new operators in the partial UV completion would, for instance, require the operator $(\p \phi)^4$ in \eqref{pUV_3} to give a dominant contribution over that from the electron loop, leading to the condition
\ba
\frac{\beta}{\Lambda^4}>\frac{1}{90(4\pi)^2\mpl^2 M^2}\,.
\ea
Hence also the irrelevant $\phi$ self-interactions have to occur at or below the scale $\Lambda=(M \mpl)^{1/2}$.

\subsection{EFT of two light scalars}

The findings of the previous sections point strongly towards the fact that scalar field theories that would otherwise be renormalizable in the absence of gravity, would be forced to carry non-renormalizable interactions at a scale much lower than the Planck scale in the presence of gravity
if we insisted on imposing the compactified positivity bounds. As another argument in favour of this finding is the example of two light scalar fields coupled to gravity and described by the action
\be
\label{eq:2fields}
\L_{\chi,\phi}=\sqrt{-g}\left[\frac{\mpl^2}{2}R-\frac{1}{2}(\partial\chi)^2-\frac{1}{2}(\partial\phi)^2-\frac{1}{2}m^2\phi^2-aM\frac{\phi^3}{3!}-\frac{\lambda}{4!}\phi^4\right]\,.
\ee
We have computed all the necessary ingredients for the $\chi\phi\to\chi\phi$ scattering in the subsections \ref{sec:tree}, \ref{sec:phi3}, \ref{sec:quartic}. All the $s^2$ contributions to the scattering amplitude were found to be negative and are thus in contradiction with the compactified positivity bounds. We have also considered possible renormalizable couplings between the two fields: $\chi\phi^2, \phi\chi^2$ in appendix \ref{app:addmore}. All of these are still negative. Similarly as in the action \eqref{pUV_3} the only positive $s^2$ contribution to the scattering amplitude can arise from the derivative interaction
\be
\frac{\delta}{\Lambda^4}(\partial_\mu\chi\partial_\nu\phi)^2\,.
\ee
Combining it with the result \eqref{amp_phi3}  from the scattering process due to the $\phi^3$ vertex and applying the positivity bounds gives the condition
\be
\frac{\delta}{\Lambda^4}\gtrsim\frac{a^2 M^2}{\mpl^2m_{\rm ph}^4}\,,
\ee
leading again to $\Lambda\lesssim(m_{\rm ph}^2 \mpl/M)^{1/2} $. For the typical case, naturalness arguments would suggest $a M \sim {\cal O}(m_{\rm ph})$ which would then imply the presence of new UV physics interactions at a scale at or below $(m_{\rm ph} \mpl)^{1/2}\ll \mpl$. For instance applying this argumentation to light models of dark matter would suggest that any such model with a dark matter mass smaller than $m_{\rm ph}< 10^{-3}$eV would also need to include interactions with all the other fields in nature, including all the other Standard Model fields at or below the  TeV scale.

\subsection{Summary of new UV physics}

To put in perspective the implications of the previous bounds, let us go back to the analogy between the model \eqref{pUV_3} and QED minimally coupled to gravity, where $M$ plays the role of the electron mass. The `need' for operators of the form $(\p \chi \p \phi)^2$ for any light (scalar) field $\chi$ even if it had no contact with $\phi$ other than gravitationally, would seem to suggest, in this analogy with QED, that photons ought to couple for instance  with dark matter and with any other light field, including themselves at or below the scale $\Lambda=(M \mpl)^{1/2}$. Taking $M$ to be the electron mass would lead to interactions at or below the scale $\Lambda\sim 10^{16}$eV. While this is a relatively high scale, it would still have profound consequences for our understanding of the Standard Model. Even more interesting, these bounds would seem to suggest that any model of say dark matter that includes cubic interactions as depicted in \eqref{eq:2fields} would also need to include additional interactions with all the fields of nature, including those of the Standard Model at scale $\Lambda=(M \mpl)^{1/2}$, where $M$ now is related to the mass of the dark matter field. This would imply interactions below the TeV scale for models of dark matter with \mbox{$M\lesssim$ meV}.
 While of course these contributions disappear in the limit $\mpl\to \infty$ where gravity decouples, such a phenomenon would require a cutoff for gravity coupled to matter well below the Planck scale, which would be a statement of unprecedented magnitude.

\section{Infrared Regulator and 3d gravity}\label{Infrared}

In the previous section we have seen that generic renormalizable field theories coupled to gravity violate the compactified positivity bounds unless new physics is introduced at the low scale $\Lambda=(M \mpl)^{1/2}$.
This is a remarkably strong conclusion and to take its implications seriously we should be clear that the assumptions taken in its derivation are valid. One concern relates to the compactification procedure and whether the limits $t \rightarrow 0$ and $L \rightarrow \infty$ can be taken to commute as noted in \cite{Loges:2019jzs}. Explicit calculations at fixed $L$ show that the positivity bounds are violated for finite nonzero $t$ and for any $L$ so this seems unlikely to be the issue. Furthermore it seems implausible that we are either not allowed to consider 3d compactifications, or that we are not allowed to consider the 3d theory \eqref{3Deff2} in its own right.
Far more likely the issue lies with the very poorly defined nature of 3d scattering amplitudes with massless gravitons \cite{tHooft:1988qqn,Ciafaloni:1992hu,Deser:1993wt,Zeni:1993ew}.

\subsection{IR regulated 3d amplitudes}

 The central issue is that in 3 dimensions, a point particle generates a conical deficit angle and as such its influence is felt at arbitrarily large distances \cite{Deser:1983tn,Deser:1988qn}. This means that the traditional notion of an $S$-matrix is poorly defined. In perturbation theory this shows up as infrared divergences that need to be regulated. One way to regulate the infrared behaviour is to add a mass to the graviton. In this section we shall focus on the effect of such a regulator on the eikonal scattering amplitude.
For instance, a static particle with mass $M_*$ can be described by the cosmic string metric with deficit angle $\delta= \frac{M_*}{2\pi M_3}$, where $M_3$ is the 3d Planck scale
\be
\d s^2 = - \d t ^2 + e^{ -2\delta \ln r} ( \d r^2 + r^2 \d \theta^2)\,,
\ee
with $\theta \in [0,2\pi]$. This can be regulated by
\be\label{cosmicstring}
\d s^2 = - \d t ^2 + e^{2\phi(r)} ( \d r^2 + r^2 \d \theta^2)\,,
\ee
for which the regulated Newtonian potential satisfies
\be
-\nabla^2_2 \phi(r)+ m_g^2 \phi(r) = \frac{M_*}{2M_3}\delta^2(x)\,,
\ee
with $m_g$ the graviton mass. The large distance Yukawa fall-off of $\phi$ renders scattering in the regulated metric \eqref{cosmicstring} well-defined.
For a massless particle of energy $\omega$ scattering in this geometry, in the eikonal limit the scattering phase shift is \cite{Ciafaloni:1992hu}
\ba\label{eikonalphase3}
\delta_{\ell}(s) &\approx& \frac{1}{8\sqrt{s}} \int_0^{2 \pi} \frac{\d \theta}{2 \pi}  e^{-i \ell \theta} \frac{su}{M_3(t-m_g^2)}\approx\frac{1}{4 s M_3} \int_{-\infty}^{\infty} \frac{\d q}{2 \pi} e^{-i q b} \frac{s^2}{q^2+m_g^2}=\frac{s}{8 M_3 m_g} e^{-|b| m_g} \nn \\
&\approx&  \frac{s}{8 M_3 m_g} -\frac{s}{8 M_3} | b | + {\cal O}(m_g)\,,
\ea
with impact parameter $b = \ell/\omega$, and with $q^2 = - t$, $s = (M_*+\omega)^2 - \omega^2 \approx 2 M_* \omega$, at high frequencies $\omega \gg M_*$. It is apparent that the phase shift is divergent as $m_g \rightarrow 0$. To define the limit $m_g \rightarrow 0$ we may rescale the scattering amplitude by an IR divergent phase (which should not change the physics)
\be\label{phaseredef}
e^{2i \delta_{\ell}(s) }= e^{i \frac{s}{4 M_3 m_g} } e^{2i \tilde \delta_{\ell}(s) }\,,
\ee
and use $\tilde \delta_{\ell}(s)$ to define a scattering amplitude which is finite in the limit $m_g \rightarrow 0$
\be \label{eikonalphase1}
\lim_{m_g \rightarrow 0} \tilde \delta_{\ell}(s) = -\frac{s}{8 M_3} | b | = - \frac{M_*}{4 M_3} |\ell| \, ,
\ee
in the physical region where $s>0$.
This matches the leading term obtained from the semiclassical conical deficit calculation of \cite{Deser:1988qn}, which gives
\be \label{eikonalphase2}
 \tilde \delta_{\ell}(s) = - \frac{\pi}{2} |\ell| \( \frac{1}{1- \frac{M_*}{2 \pi M_3}}-1\) \, .
\ee
Note however that the phase shift redefinition \eqref{phaseredef} is far from innocent. The positivity bound arguments rely on the assumption that the fixed $t$ scattering amplitude respects the Jin-Martin bound $\lim_{|s| \rightarrow \infty} s^{-2} \A(s,t)=0$ for all complex $s$ on the first Riemann sheet. If this were not the case, we could not have written the assumed dispersion relation. When there is a mass gap, and for $m_g>0$ we expect the 3d Froissart bound to hold \cite{Chaichian:1992hq}. However the redefinition \eqref{phaseredef} factorizes out an entire function that is not polynomially bounded, and grows exponentially in the lower half of the complex $s$ plane, undermining the boundedness assumptions. We thus have no reason to expect that the regulated 3d scattering amplitudes for $m_g=0$ respect any polynomial boundedness.

\subsection{Unitarity in 3d}

Even more damning though is the failure for the scattering amplitude to have the assumed analyticity structure, positivity and smoothness properties. By assumption, if the scattering amplitude were to satisfy $\lim_{|s| \rightarrow \infty} s^{-2} \mathcal A(s,t) =0$ for fixed $t$ with only the usual poles and branch cuts, then up to the unknown subtraction constants and known poles, it should be entirely determined by its imaginary part $\text{Im}\,\mathcal A(s,t)$ for $s \ge 0$.
From the 3d partial wave expansion in the physical region, we have
\be\label{partialwave}
\mathcal A(s,t) = \frac{ 4\sqrt{s}}{i} \sum_{\ell=-\infty}^{\infty} e^{i \ell \theta} \( e^{2i \delta_{\ell}(s)} -1 \) \, .
\ee
The imaginary part (given $\delta_{\ell}(s)=\delta_{-\ell}(s)$ and splitting the phase up to its real and absorptive parts $\delta_{\ell}(s) = \delta^r_\ell(s) + i \delta^a_\ell(s)$) is
\be
\text{Im}\,\mathcal A(s,t) = 4 \sqrt{s} \left[\(1- \cos(2 \delta^r_0(s)) e^{-2 \delta^a_0(s)}\) +  2 \sum_{\ell=1}^{\infty}\cos(\ell \theta) \(1- \cos(2 \delta^r_{\ell}(s)) e^{-2 \delta^a_\ell(s)}\)   \right] \, .
\ee
In particular for $t=0$
\be\label{imaginaryforward10}
\text{Im}\,\mathcal A(s,0) = 4 \sqrt{s} \left[\(1- \cos(2 \delta^r_0(s)) e^{-2 \delta^a_0(s)}\) +  2 \sum_{\ell=1}^{\infty} \(1- \cos(2 \delta^r_{\ell}(s) )e^{-2 \delta^a_\ell(s)}\)   \right] >0\, ,\ee
is obviously positive if the sum is convergent, as are $\partial_t^N \text{Im}\,\mathcal A(s,0) $.
Assuming the validity of the eikonal result for $m_g \to 0$ \eqref{eikonalphase1} at large $\ell$, or at least that $\lim_{|\ell| \rightarrow \infty} \delta_{\ell}(s) = - \alpha(s) | \ell|$, this series diverges. In the IR regulated case \eqref{eikonalphase3} it is by contrast convergent for fixed $s$ as the graviton mass enforces an effective maximum angular momentum ${\ell}_{\rm max} \sim \frac{s}{2 m_g M_*}$.

By contrast the usual implicit approach in the massless case \cite{Deser:1988qn,Ciafaloni:1992hu} is to enforce convergence with an implicit $i \epsilon$ regulator, take
\be
\text{Im}\,\mathcal A(s,t) = 4 \sqrt{s} \left[\(1- \cos(2 \delta^r_0(s))  e^{-2 \delta^a_0(s)}\) +  2 \sum_{\ell=1}^{\infty}\cos(\ell \theta)e^{- \epsilon |\ell|} \(1- \cos(2 \delta^r_{\ell}(s)) e^{-2 \delta^a_\ell(s) }\)   \right] \, .
\ee
This is qualitatively similar to the regulated case with $\epsilon \sim 1/{\ell}_{\rm max}$, the difference being that the regulated problem enforces convergence by having $\delta_{\ell}(s) \rightarrow 0$ for $\ell > {\ell}_{\rm max}$. For a phase shift of the form $\delta_{\ell}(s) = - \alpha(s) |\ell|$ the $i \epsilon $ regulated expression gives the imaginary part as a sum over delta functions.
\ba
\text{Im}\,\mathcal A(s,t) &\sim&  8 \sqrt{s}  \sum_{\ell=1}^{\infty}\cos(\ell \theta)e^{- \epsilon |\ell|} \(1- \cos(2 \alpha |\ell|)   \)    \,  \\
&=&4 \pi \sqrt{s}   \left[2\tilde \delta(\theta) - \tilde \delta(\theta+ 2 \alpha(s)) - \tilde \delta(\theta-2 \alpha(s))  \right]\,, \label{3deltas}
\ea
where the $\tilde \delta(\theta)$ denote the regulated periodic delta functions
\be
\tilde \delta(\theta) \equiv \frac{1}{2\pi} \(\frac{1}{1-e^{i \theta} e^{-\epsilon}} +\frac{1}{1-e^{-i \theta} e^{-\epsilon}} -1\) =\sum_n \delta(\theta + 2\pi n) \, ,
\ee
with the sum over $n$  ensuring angular periodicity \cite{Deser:1988qn}. Evaluating in the forward scattering limit this expression is divergent
\be\label{imaginaryforward}
\text{Im}\,\mathcal A(s,0) = 8 \pi \sqrt{s}\, \tilde \delta(0) \sim 8  \sqrt{s}\, \epsilon^{-1} \, .
\ee
As noted in \cite{Deser:1988qn} the optical theorem is no longer strictly valid due to the lack of smoothness of these functions.
Note that the first delta function $ \tilde \delta(\theta) =\sum_{n}  \delta(\theta+ 2 \pi n)$ arising in \eqref{3deltas} is not the same one that arises from the no-scattering process $\hat 1$ in the $S$-matrix split $\hat S = \hat 1 + i \hat T$ since this has already been subtracted out in \eqref{partialwave}. It is a direct reflection of the ill-defined nature of the scattering amplitude in 3d with massless gravitons due to the absence of clearly defined asymptotic states. By contrast in the IR regulated case the imaginary part of the forward amplitude \eqref{imaginaryforward10} is finite, and only diverges as $m_g \rightarrow 0$.

\subsection{Failure of positivity in 3d}

It is the previous lack of smoothness for the amplitude $\mathcal A(s,t)$ near $t=0$ that is the root of the failure of positivity in 3d as we now show. Given the divergence of the imaginary part, we could just take the perspective of simply discarding the contribution from the delta function $\tilde \delta(\theta)= \sum  \delta(\theta+ 2 \pi n)$ from \eqref{3deltas} leaving the other two $\tilde \delta$'s, which would amount to working with an amplitude in the form
\be\label{partialwave2}
\hat{\mathcal A}(s,t) = \frac{ 4\sqrt{s}}{i} \sum_{\ell=-\infty}^{\infty} e^{i \ell \theta} e^{2i \delta_{\ell}(s)}e^{- \epsilon |\ell|}   \, .
\ee
This is in fact the approach of \cite{Ciafaloni:1992hu}. Again taking $\delta_{\ell}(s) = - \alpha(s) |\ell|$  then explicitly evaluating the sum in the physical region gives
\be\label{eikonal101}
\hat{\mathcal A}(s,t)=  4 \( \frac{s \sqrt{s}  \sin(2 \alpha(s) - i \epsilon)}{s \cos(2 \alpha(s) - i \epsilon) - s- 2t}\)\,.
\ee
Most importantly we see that this expression is finite in the forward limit confirming the removal of the $t$-channel pole. The eikonal approximation for the scattering of two massless particles gives $\delta_{\ell}(s) = - \alpha(s) |\ell|$ with $\alpha(s) =\beta \sqrt{s}$ with $\beta = 1/(4 M_3)$ \cite{Ciafaloni:1992hu}. In the forward limit $t=0$, the expression \eqref{eikonal101} becomes
\be\label{eikonal102}
\hat{\mathcal A}(s,0)= -4 \( \frac{ \sqrt{s}  \sin(2 \beta \sqrt{s}- i \epsilon)}{1-\cos(2 \beta \sqrt{s}- i \epsilon) }\) = - 4 \sqrt{s} \cot( \beta \sqrt{s}- i \epsilon) \, .
\ee
The $s$-poles are at the locations $2 \beta \sqrt{s}= 2 |n| \pi$, and their imaginary parts are consistent with  \eqref{3deltas}
\ba
\text{Im}\,\hat{\mathcal A}(s,0) &=& 4 \pi \sqrt{s}  \sum_{n=-\infty}^{\infty} \left[- \delta(2 n \pi+ 2 \beta \sqrt{s}) -  \delta(2 n \pi-2 \beta \sqrt{s})  \right] \, \\
&=& - 8 \pi   \sum_{n=0}^{\infty} \frac{s}{\beta} \delta( s - n^2 \pi^2/\beta^2)\,.
\ea
However, it is now apparent that $\text{Im}\,\hat{\mathcal A}(s,0)<0$ and so positivity is lost. Indeed more generally working with $\hat{\mathcal A}(s,t)$ we have
\be\label{positivitylost}
\text{Im}\,\hat{\mathcal A}(s,t) = -4 \sqrt{s} \left[\cos(2 \delta^r_0(s)) e^{-2 \delta^a_0(s)} +  2 \sum_{\ell=0}^{\infty}\cos(\ell \theta)  \cos(2 \delta^r_{\ell}(s)) e^{-2 \delta^a_\ell(s)}e^{- \epsilon |\ell|}  \right] \, ,
\ee
which is sign indefinite even in the forward limit, and so destroys the crucial positivity property utilized in the dispersion relation arguments. Unitarity is of course still intact $|e^{2i \delta_{\ell}(s)}| \le 1$, but its implication for the positivity of the dispersion relation is different. \\

The key point is that the standard statement of unitary rests on the decomposition $\hat S = \hat 1 + i \hat T$, so that $-i (\hat T - \hat T^{\dagger} ) = \hat T \hat T^{\dagger}$, which relies on the notion that there is some probability for no scattering. However in 3d massless gravity, it would be impossible to have a scenario where no scattering occurs since each mass distorts the metric at infinity by means of a deficit angle. This is why if we try to enforce the split $\hat S = \hat 1 + i \hat T$ as in \eqref{partialwave}, then the resulting scattering amplitude will result in a delta function that compensates the $\hat 1$ as in \eqref{3deltas}. The amplitude which is well-defined is $\hat {\cal A} $ which follows directly from $-i \hat S$, but this does not have a positive imaginary part since unitarity\footnote{It has been suggested that this may be resolved by working with a redressed scattering amplitude for which the gravitational interactions are removed \cite{PrivateDiscussion}. However, whilst such a redressed  amplitude would certainly be unitarity, there is to date no explicit construction that is crossing symmetric and analytic which are crucial ingredients in the derivation of the positivity bounds.} is realized through $(-i \hat S)(-i \hat S)^{\dagger} = \hat 1$.

\subsection{Analytic structure of the amplitude in 3d}

A closer inspection of \eqref{eikonal102} shows that it has the wrong analytic structure. If we attempt to construct the function from its imaginary part via a dispersion integral we obtain rather
\be
-\frac{4}{\beta}+\frac{s}{\pi} \int_0^{\infty} \d s' \frac{\text{Im}\,\hat{\mathcal A}(s',0)}{s'(s'-s-i \epsilon)} =-\frac{4}{\beta} - 8    \sum_{n=0}^{\infty} \frac{1}{\beta} \frac{s}{(\frac{n^2 \pi^2}{\beta^2}-s-i \epsilon)}  = 4 \sqrt{s} \cot( \beta \sqrt{s}+i \epsilon) \, ,
\ee
which has the same imaginary part but opposite sign real part in the physical region. The reason for this failure can be traced to the eikonal expression for the partial wave $S$-matrix $e^{2i \delta_{\ell}(s)} = e^{-2 i \beta \sqrt{s} |\ell|}$. Analytically continuing into the Euclidean region via a counterclockwise rotation $s= e^{i \pi} s_E$ gives the exponentially growing behaviour
$e^{2i \delta_{\ell}(s)} = e^{2  \beta \sqrt{s_E} |\ell|}$ for which the sum over $\ell$ does not converge\footnote{By contrast, if $\beta$ has been of opposite sign $\beta=-\beta'$ with $\beta'>0$, then \eqref{eikonal101}  would have been
\be \label{eikonal103}
\hat{\mathcal A}'(s,t)= - 4 \( \frac{s \sqrt{s}  \sin(2 \beta' \sqrt{s} + i \epsilon)}{s \cos(2 \beta' \sqrt{s} + i \epsilon) - s- 2t}\)\,,
\ee
which has the correct analytic structure and for which the bounded partial waves would be $e^{2i \delta_{\ell}(s)} = e^{-2  \beta' \sqrt{s_E} |\ell|}$.}. This failure of analyticity can be attributed to the failure of the eikonal approximation for the phase shift. This can be solved by a variant application of the eikonal approximation for which the sum \eqref{partialwave2} is replaced by an integral over $\ell$, giving for small $t$
\be
\hat{\mathcal A}(s,t) \sim   \frac{4 \beta s^2}{-t-\beta^2 (s-i \epsilon)^2} \, .
\ee
Expressed in a crossing symmetric form (one of many) \cite{Ciafaloni:1992hu}, a suitable ansatz is
\be\label{eikonal104}
\hat{\mathcal A}(s,t) =   \frac{4 \beta su}{t-\beta^2 s u} =a(t) + \frac{b(t)}{\mu(t)-s} + \frac{b(t)}{\mu(t)-u} \, ,
\ee
with $b(t)= 2 q^2/\sqrt{q^2+\frac{\beta^2}{4}q^4}$, $\mu(t)=\frac{1}{2} q^2+ \frac{1}{\beta} \sqrt{q^2+\frac{\beta^2}{4}q^4}$, $a(t)=-\frac{4}{\beta}$ and   with $q^2=-t$, we can see that
these expressions amount to replacing the naive eikonal phase shift $e^{2 i \delta_{\ell}(s)}  = e^{-2 i \beta |\ell| \sqrt{s}}$ with
\be
e^{2 i \delta_{0}(s)} =\frac{1- i \beta \sqrt{s}}{(1+\beta^2 s)} \, , \quad \text{ and }\quad e^{2 i \delta_{\ell}(s)} = \frac{1}{(1+\beta^2 s)} \( \frac{1- i \beta \sqrt{s}}{1+ i \beta \sqrt{s}}\)^{|\ell|} \, , \quad |\ell| \ge 1\\
 \, ,
\ee
which are bounded analytic functions up to a right-hand branch cut and multiple poles at $s= - \beta^{-2}$.  \\

The expression \eqref{eikonal104} is a crossing symmetric, analytic function of $s$ for fixed $t \le 0$ up to poles at fixed $t\le 0$ with positive imaginary parts $\Im \,\hat{\mathcal A}(s,t)>0$ for $t \le 0$. It would then appear to satisfy everything we desire for a 3d scattering amplitude. Nevertheless this positivity of the imaginary part is accidental rather than implied by unitarity due to \eqref{positivitylost}.
This is most apparent from the fact that the imaginary part vanishes in the forward limit
\be
\Im \, \hat{\mathcal A}(s,0) =\pi b(0) \delta(s- \mu(0))=0 \, ,
\ee
which follows since
\be
\frac{4 \sqrt{s}}{i}\sum_{\ell=-\infty}^{\infty} e^{2 i \delta_{\ell}} =\frac{4 \sqrt{s}}{i} \[ \frac{1- i \beta \sqrt{s}}{(1+\beta^2 s)} +2 \sum_{\ell=1}^{\infty}\frac{1}{(1+\beta^2 s)} \( \frac{1- i \beta \sqrt{s}}{1+ i \beta \sqrt{s}}\)^{\ell} \]= -\frac{4}{\beta } \, ,
\ee
is purely real even though every partial wave contributes with  a non-zero imaginary part. Furthermore the imaginary part of the fixed angle $\theta$ amplitude \eqref{eikonal104} is negative. Even if the expression \eqref{eikonal104} is used within the context of improved positivity bounds to remove the eikonal contribution, there is no reason to expect that the remaining imaginary part would be positive, due the lack of positivity of \eqref{positivitylost}. We thus conclude that there is no form of 3d scattering amplitude which is both smooth or at least finite at $t=0$ and for which positivity of its imaginary part is guaranteed to hold. Since both these properties are needed together for the derivation of positivity bounds we conclude that they do not apply in 3 dimensions.

\subsection{Perturbative expansion of 3D amplitude}

In order to make clear that the crossing symmetric ansatz $\hat{\mathcal A}(s,t) =4 \beta su/(t-\beta^2 s u) $ provided in \cite{Ciafaloni:1992hu} is the $S$-matrix and not $T$-matrix element, it is helpful to compare with its perturbative expansion in powers of inverse Planck mass. Since $\beta \sim 1/M_3$ we have for $t \neq 0$,
\be
\hat{\mathcal A}(s,t) = \frac{4 \beta su}{t} +  \frac{4 \beta^3 s^2u^2}{t^2} + \dots\,.
\ee
This matches the form of the terms we expect from a perturbative expansion of the transition matrix, but the expansion is  ill defined near $t=0$. To correctly identify the forward limit delta function we denote $t=-q^2$ and consider the integral for fixed $s$
\be
\int_{-\infty}^{\infty} \d q \, \hat{\mathcal A}(s,-q^2) =\int_{-\infty}^{\infty} \d q  \frac{4 \beta s (s-q^2)}{q^2-\beta^2 s (s-q^2)} =\int_{-\infty}^{\infty} \d q'  \frac{4 s (s -\beta^2{q'}^2)}{{q'}^2-s (s-\beta^2  {q'}^2)} \, .
\ee
Taking the limit in which gravity decouples we have
\be
\lim_{M_3 \rightarrow \infty} \int_{-\infty}^{\infty} \d q \, \hat{\mathcal A}(s,-q^2) =\int_{-\infty}^{\infty} \d q'  \frac{4 s^2}{{q'}^2-(s- i \epsilon)^2} = -4 i\pi s\, .
\ee
Then combined with the fact that $\lim_{M_3 \rightarrow \infty} \hat{\mathcal A}(s,-q^2) =0$ for $t \neq 0$, this implies $\lim_{M_3 \rightarrow \infty} \hat{\mathcal A}(s,-q^2) =-4 i\pi s \delta(q)$, confirming that $\hat{\mathcal A}(s,t) $ is indeed the full $S$-matrix amplitude.

\subsection{Positivity recovered with a mass gap}
\label{sec:massGap}

If, as implied in the previous subsections, it is indeed correct that the problems associated  with the compactified positivity derivation are the ill-defined IR behaviour of the amplitude ${\cal A}$ (as defined in \eqref{partialwave}) and the associated lack of positivity of the amplitude $\hat{\cal A}$ (as is apparent in \eqref{positivitylost}), then these issues would be resolved if the spin-2 states were massive, for which we may return to using ${\cal A}$. As we shall see, this is indeed the case and the positivity bounds are only problematic in the case of a massless graviton exchange, not for a massive one.

Returning to the four dimensional amplitude, had we considered the exchange of a massive spin-2 field of mass $m_g$, the corresponding result for the scattering amplitude would be instead
\be\label{amp_pUV_m}
\mathcal A_{\rm pUVm}(s,t)=-\frac{s^2}{\mpl^2(t-m_g^2)}-\frac{1}{90(4\pi)^2}\frac{\alpha^2s^2t}{M^2\mpl^2(t-m_g^2)}+\mathcal O((t-m_g^2)^0)\,.
\ee
As expected, the only relevant effect is the shift of the $t$-channel pole away from the origin, while the overall sign remains unaffected. In particular we emphasize that the slight modification of the polarization structure of the massive vs massless spin-2 propagator bears no consequences to this discussion. \\

The presence of a mass gap allows us to safely continue the partial wave expansion from the physical region $t \le 0$ to the region $0\le t<m_g^2$  without fear of admonition. The positivity bounds can then be applied for any $t$ in the region $0\le t<m_g^2$ \cite{deRham:2017avq}. Remarkably, the amplitude  $\mathcal A_{\rm pUVm}$ manifestly satisfies the positivity bounds as expected.  To see this we denote $t = x m_g^2 + \tau$ with $0<x<1$ and expand in $\tau$
\ba\label{amp_pUV_m_2}
\mathcal A_{\rm pUVm}(s,t)&=&\frac{s^2}{\mpl^2((1-x) m_g^2- \tau)}+\frac{1}{90(4\pi)^2}\frac{\alpha^2s^2 (x m_g^2 + \tau)}{M^2\mpl^2((1-x) m_g^2- \tau)}+\mathcal O((t-m_g^2)^0)\, , \nn \\
&=& \frac{s^2}{\mpl^2(1-x) m_g^2} +\frac{1}{90(4\pi)^2}\frac{\alpha^2 s^2 x}{M^2\mpl^2(1-x)}+\sum_{n=1}^{\infty} c_n(s) \tau^n \, .
\ea
All coefficients of the expansion are manifestly positive $c_n(s)>0$ as required, and furthermore
\be
\frac{\partial^2 {\cal A}_{\rm pUVm}(s,\tau)}{\partial s^2}\Big|_{ \tau=0,  \, 0<x<1} >0 \, ,
\ee
without any assumption on the magnitude of $\alpha$. This strongly suggests that any issue with satisfying the positivity bounds in the massless case is not an issue with the partial UV theory considered in this work, rather, it suggests that the contradiction apparent in~\eqref{contradiction} with the compactified positivity bounds with a massless graviton is due to the absence of a mass gap in the massless case and the associated poor IR behaviour. The fact that positivity bounds are applicable for massive spin-2 states has been used extensively in recent works \cite{Cheung:2016yqr,Bonifacio:2016wcb,deRham:2017xox,Bellazzini:2017fep,deRham:2018qqo,Alberte:2019lnd,Alberte:2019xfh,Alberte:2019zhd}.


\section{Conclusions}\label{sec:conclusions}

Positivity bounds are expected to apply to gravitational theories whenever there exists a clean decoupling limit $\mpl \rightarrow \infty$ for which the graviton decouples from other degrees of freedom. This is the case whenever the low-energy scattering amplitude takes the schematic form
\be\label{scaling1}
{\cal A}(s,t) \sim - \frac{s^2}{\mpl^2 t} + \frac{c}{M^4}s^2+ \dots \, ,
\ee
with $c \sim {\cal O}(1)$ for which we may scale $\mpl \rightarrow \infty$ for fixed $M$ so that the non-gravitational positivity bounds imply $c>0$.
Related arguments have been given in Ref.~\cite{Hamada:2018dde}, where it is argued that positivity bounds should apply for tree-level UV completions when the higher spin states Reggeizing graviton exchange are subdominant in the matter (e.g. photon) scattering.\footnote{A weakly--coupled tree-level UV completion is often assumed ``as a safety net" for practical computational purposes in applying the positivity bounds so as to ensure that amplitudes are dominated by tree-level diagrams, however the positivity bounds themselves  as expressed for instance in \eqref{positivity} are derived with no prior limitation on the type of UV realization, so long as it is a standard one as far locality, unitarity, Lorentz invariance and causality are concerned.}
In the examples discussed here, the scattering amplitude rather takes the form
\be\label{scaling2}
{\cal A}(s,t) \sim - \frac{s^2}{\mpl^2 t} + \frac{\tilde c}{M^2 \mpl^2}s^2+ \dots \, ,
\ee
with $\tilde c \sim {\cal O}(1)$. In this situation we can no longer decouple gravity without making the whole effect vanish (or bringing the cutoff to zero). Furthermore in all of our examples, $\tilde c$ arises from loop effects rather than tree-level/higher spin UV physics and so is not covered by the argument of \cite{Hamada:2018dde}. It is thus no longer clear whether we require $\tilde c>0$. As discussed in  \cite{deRham:2020zyh}, for the scaling \eqref{scaling1}, having $c \sim {\cal O}(1)<0$ clearly leads to superluminal propagation and violation of causality, which is consistent with previous expectations on the connection between positivity bounds  \cite{Pham:1985cr,Ananthanarayan:1994hf} and causality \cite{Adams:2006sv}. By contrast for the scaling \eqref{scaling2}, $\tilde c \sim {\cal O}(1)<0$ does not lead to any resolvable violation of microcausality \cite{deRham:2020zyh,Hollowood:2015elj}. Consequently we can no longer rely on causality arguments to argue for any bound on $\tilde c$. The proposed compactified positivity bounds of \cite{Bellazzini:2019xts} attempt to bypass this by using positivity of 3d scattering amplitudes to indirectly infer $\tilde c>0$ even for the scaling choice \eqref{scaling2}. If true, these would have profound consequences, most notably for the weak gravity conjecture \cite{ArkaniHamed:2006dz,Cheung:2014ega,Hamada:2018dde,Bellazzini:2019xts}. \\

In the present article, we have shown that these proposed compactified positivity bounds are generically violated for typical renormalizable theories coupled to gravity unless new physics is introduced at the parametrically low scale $\Lambda \sim (M \mpl)^{1/2}$. One may take the perspective that this simply implies that our renormalizable theory lies in the swampland, however this result remains relatively stable under relevant and marginal deformations, and similar observations hold for the more realistic case of QED \cite{newpaper}. Such a conclusion about the low scale $\Lambda \sim (M \mpl)^{1/2}$ would have profound implications for our understanding of the landscape of theories with consistent Lorentz invariant, analytic, UV completions.\\

There are however as discussed in section~\ref{Infrared} a number of technical issues with the derivation of the compactified positivity bounds proposed in \cite{Bellazzini:2019xts} that prevents us from immediately accepting these conclusions. Most critical is the fact that 3d scattering amplitudes in the presence of massless spin-2 particles are poorly defined, not least because gravitational interactions do not vanish at infinity. This shows up as singular behaviour in the usual definition of the scattering amplitude in the forward limit. We show explicitly that attempting to remove this singular behaviour gives a scattering amplitude that does respect analyticity but no longer respects positivity of its imaginary part (despite unitarity being intact). This undermines the applicability of any positivity bound to 3d gravitational scattering amplitudes (at least for massless gravitons) and shows how the issues with the 4d massless $t$-channel pole manifest themselves through slightly different but ultimately equivalent pathologies in 3d.  \\

{\bf Conjecture:}  We conclude by postulating a conjecture on the implications of the $t$--channel pole subtracted positivity bounds and the amount by which they may in principle be violated assuming that the arguments of \cite{Bellazzini:2019xts} are indeed flawed. Given a theory with a scattering amplitude with low-energy expansion of the form \eqref{scaling1} where $M$ is the cutoff of the low-energy EFT, we can at most expect a bound in the weak sense\footnote{A similar conclusion was pointed out in the latest arXiv version of \cite{Bellazzini:2019xts} assuming an exact $s^2$ UV behaviour of the amplitude. We make here no such assumptions in the derivation of this result. Rather this conjecture is here tied to the requirement of causality as emphasized in \cite{deRham:2019ctd} and \cite{deRham:2020zyh}.}
\be\label{eq:OMpl2}
c > - \frac{M^2}{\mpl^2} \times {\cal O}(1) \, .
\ee
Generically this seems to suggest that even though negative coefficients could in principle be compatible with standard high-energy completion, they ought to be highly suppressed, and the scaling with $\mpl$ is such that one recovers the standard positivity bounds $c>0$ whenever a decoupling limit $\mpl\to  \infty$ can be taken. Even if suppressed, allowing for the very possibility of having a small negative coefficient would have important implications for the weak gravity conjecture where one of its manifestations relies precisely on operators suppressed with the precise same powers of $\mpl$, \cite{Cheung:2014ega,Hamada:2018dde}.\\

As mentioned previously, it is likely that the bound \eqref{eq:OMpl2} can be further refined to $c>0$ using appropriate scaling arguments for a restricted class of UV completions, for instance for tree level higher spin/Regge state UV completions with subdominant couplings to matter as argued in \cite{Hamada:2018dde}. Interestingly, this would imply that observing a negative coefficient $c$ experimentally  could be interpreted as indications against these types of completions.\\

The scaling of the bound \eqref{eq:OMpl2} is similar to that found in \cite{deRham:2020zyh} where it was argued that superluminalities within low-energy gravitational theories are consistent with causality and can emerge from standard and causal high-energy completions so long as the amount of superluminality scales similarly at least as $\mpl^{-2}$ and vanishes in a decoupling limit where gravity decouples for which traditional causality arguments apply. In the specific case of scalar QED this is consistent with the observations of \cite{Hollowood:2015elj}. Future work is needed to better understand the role of both causality and positivity in gravitational effective theories.

\bigskip
\noindent{\textbf{Acknowledgments:}}
We would like to thank Brando Bellazzini, Andrey Khmelnitsky,  Matthew Lewandowski, Arshia Momeni, Justinas Rumbutis and  Javi Serra for useful discussions.
 The work of AJT and CdR is supported by STFC grants ST/P000762/1 and ST/T000791/1. CdR thanks the Royal Society for support at ICL through a Wolfson Research Merit Award. LA and CdR are supported by the European Union's Horizon 2020 Research Council grant 724659 MassiveCosmo ERC--2016--COG. CdR is also supported by a Simons Foundation award ID 555326 under the Simons Foundation's Origins of the Universe initiative, `\textit{Cosmology Beyond Einstein's Theory}' and by a Simons Investigator award 690508.  SJ is supported by an STFC studentship. AJT thanks the Royal Society for support at ICL through a Wolfson Research Merit Award.

\appendix
\section{One--loop electron contribution to photon scattering in scalar QED}\label{app:box}

In this appendix we shall provide the key steps for determining  the one--loop electron contribution to the $\phi\phi\to\phi\phi$ scattering amplitude as illustrated in Fig.~\ref{fig:psibox}.
Accounting for the three channels, the amplitude is given by
\be
\mathcal A_{\psi \rm -box}=\mathcal A_{\psi \rm -box}(s,t,u)+\mathcal A_{\psi \rm -box}(s,u,t)+\mathcal A_{\psi \rm -box}(t,u,s)\,,
\ee
where the individual amplitudes read
\be
\mathcal A_{\psi \rm -box}(s,t,u)=\frac{\alpha^4M^4}{\pi^2}\int_{\mathcal R}\text{d}x\,\text{d}y\,\text{d}z\,\frac{1}{\Delta^2}\,,
\ee
and the denominator is given by $\Delta=M^2+sz(x+y+z-1)-txy$.
The integration region $\mathcal R$ is defined as $\mathcal R= \{x,y,z\in\mathbb R\,|\,x,y,z\geq0, x+y+z\leq 1\}$. Going to the forward limit ($t=0$) we can evaluate this amplitude in an expansion of $\frac{s^2}{M^4}$. We find
\be
\mathcal A_{\psi \rm -box}=\frac{\alpha^4}{2\pi^2}\left(1+\frac{s^2}{60M^4}\right)+\mathcal O\left(\frac{s^4}{M^8}\right)\,.
\ee
Matching these contributions to contact interactions in the one-loop effective action leads to the following operators to be included in \eqref{IR_0}:
\be
\mathcal L_{\rm IR}\supset\sqrt{-g}\left[\frac{\alpha^4}{2\pi^2}\frac{\phi^4}{4!}+\frac{1}{480\pi^2}\frac{\alpha^4}{M^4}\left(\partial\phi \right)^4\right]\,.
\ee

\section{Amplitudes from the Perspective of the Partial UV theory}\label{sec:uv_ampl}
\begin{figure}
\centering
\begin{subfigure}{0.3\textwidth}
\centering
\includegraphics[height=1.5in]{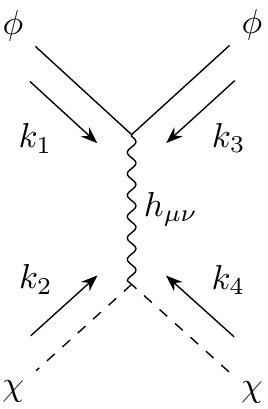}
\caption{Tree diagram for $\mathcal A_{\text{tree,0}}$.} \label{fig:1a}
\end{subfigure}
\hspace*{\fill} 
\begin{subfigure}{0.3\textwidth}
\centering
\includegraphics[height=1.5in]{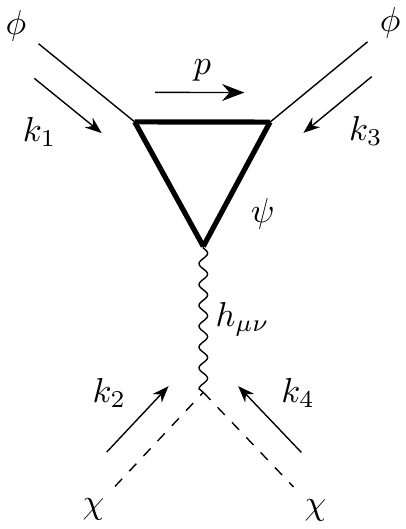}
\caption{One-loop diagram for $\mathcal A_{\phi\psi^2}$.} \label{fig:1b}
\end{subfigure}
\hspace*{\fill} 
\begin{subfigure}{0.3\textwidth}
\centering
\includegraphics[height=1.5in]{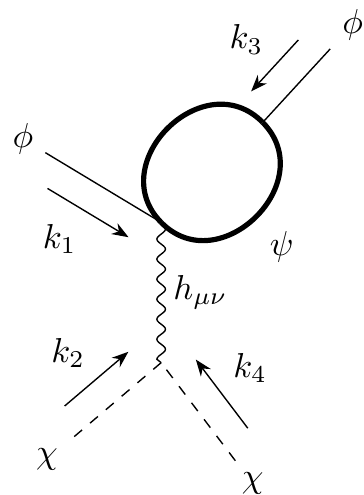}
\caption{One-loop diagram for $\mathcal A_{\phi\psi^2h}$.} \label{fig:1c}
\end{subfigure}
\caption{The detailed Feynman diagrams for the three scattering processes contributing to the $\chi\phi\to\chi\phi$ one-loop scattering in the UV theory \eqref{pUV_0}. The wiggly lines correspond to graviton propagators, while  bold lines stand for  propagators of the heavy field $\psi$.} \label{fig:1}
\end{figure}

In this section we give the details for the computations of the $\chi\phi\to\chi\phi$ scattering amplitude in the UV theory \eqref{pUV_0}. The relevant Feynman diagrams are shown in Fig.~\ref{fig:1}. We introduce the following short-hand notations for the momentum-dependent vertices
\begin{align}\label{rules}
\begin{split}
\vcenter{\hbox{\includegraphics[width=2cm]{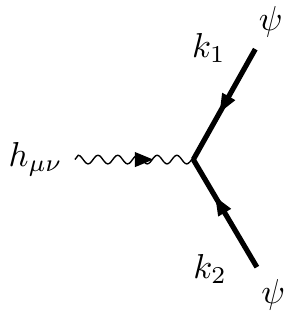}}}\qquad &V_M^{\mu\nu}(k_1,k_2)\equiv\frac{-i}{2\mpl}\bigg(k_1^\mu k_2^\nu+k_1^\nu k_2^\mu-\eta^{\mu\nu}\left(k_1\cdot k_2-M^2\right)\bigg)\,,\\
\vcenter{\hbox{\includegraphics[width=2cm]{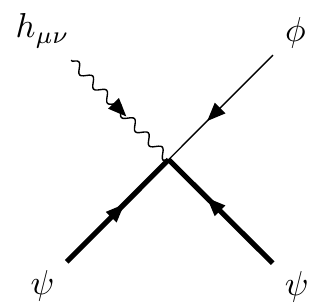}}} \qquad&V^{\mu\nu}_{\phi\psi^2h}\equiv\frac{-i\alpha M}{\mpl}\eta^{\mu\nu}\,,\\
\vcenter{\hbox{\includegraphics[width=2cm]{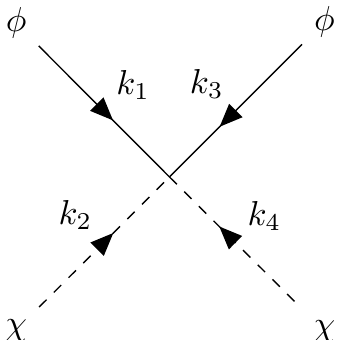}}} \qquad&V_{\phi^2\chi^2}\equiv\frac{2iC\alpha^2 }{M^2\mpl^2}\bigg((k_1\cdot k_4)(k_2\cdot k_3)+(k_1\cdot k_2)(k_3\cdot k_4)\bigg)\,,
\end{split}
\end{align}
where we use the $(-,+,+,+)$ signature and all the momenta are taken to be ingoing. We also denote the scalar field propagators by $\Delta_\phi(p)$, $\Delta_\psi(p)$, $\Delta_\chi(p)$ and assign $D_{\mu\nu;\alpha\beta}(p)$ to the graviton propagator in the harmonic gauge; the $\phi\psi^2$ interaction vertex is denoted by $V_{\phi\psi^2}\equiv-2i\alpha M$. Finally, to evaluate the loop integrals we use the standard dimensional regulation procedure that for $d$ spacetime dimensions reads
\be\label{dimreg}
\mu^\epsilon\int \frac{\text{d}^{d}k}{(2\pi)^{d}}\frac{(k^{2})^{p}}{(k^{2}+\Delta- i \varepsilon)^{n}}=\mu^\epsilon\frac{i\Gamma\left(\frac{d}{2}+p\right)\Gamma\left(n-p-\frac{d}{2}\right)}{(4\pi)^{d/2}\Gamma\left(\frac{d}{2}\right)\Gamma\left(n\right)}(\Delta- i \varepsilon)^{\frac{d}{2}+p-n}\equiv\mathcal{I}^{p}_{n}\,.
\ee

\subsection{Tree--level}\label{sec:tree}
The amplitude of the tree-level diagram in Fig.~\ref{fig:1a} is found to be (using the notations of \eqref{rules})
\be\label{tree0}
\begin{split}
i\mathcal A_{\text{tree,0}}&=V^{\mu\nu}_{m}(k_1,k_3)D_{\mu\nu;\alpha\beta}(k_1+k_3)V^{\alpha\beta}_{0}(k_2,k_4)\\
&=\frac{-i}{\mpl^2}\frac{1}{t}s(s+t)=\frac{i}{\mpl^2t}su\,,
\end{split}
\ee
where we have used the Mandelstam variables
\be
s=-(k_{1}+k_{2})^{2}\,,\quad t=-(k_{1}+k_{3})^{2}\,,\quad u=-(k_{1}+k_{4})^{2}\,,
\ee
with $k_i^2=0$ for $i=1,2,3,4$. This result is independent on the spacetime dimension and shows the presence of a $t$-channel pole in the tree-level scattering of two scalar fields via a massless graviton exchange.

Before we set off computing the one-loop amplitudes let us note that the $-\alpha M\phi\psi^2$ interaction introduces a shift in the self-energy of $\phi$ in the resummed propagator and leads to wavefunction renormalization $Z_\phi$  defined as
\be
\frac{-i}{p^2+m^2+\Sigma(p)}=\frac{-iZ_\phi}{p^2+m^2_{\text{phys}}}\,,
\ee
with the physical mass for $\phi$ determined from the relation
\be\label{phys_mass}
\left[p^2+m^2+\Sigma(p)\right]_{p^2=-m_{\text{phys}}^2}=0\,,
\ee
where in this case the bare mass $m=0$.
In particular, the shift in self-energy arising from the $\psi$ loop to quadratic order in $\alpha$ is found as:
\be
\begin{split}
-i\Sigma(p)=\vcenter{\hbox{\includegraphics[width=3cm]{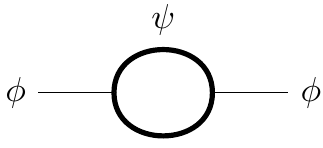}}}&=\frac{4\alpha^2M^2}{2}\int \frac{\d^dk}{(2\pi)^d}\frac{1}{[k^2+M^2]}\frac{1}{[(k-p)^2+M^2]}\\
&=\frac{i\alpha^2M^2}{8\pi^2}\int_0^1\text{d}x\,\left[\msbar+\log\frac{\mu^2}{M^2+x(1-x)p^2}\right]\,,
\end{split}
\ee
where the $1/2$ on the first line is the symmetry factor of the diagram and the latter equality is obtained for $d=4-\epsilon$. Here we denote the standard minimal subtraction scheme ($\msbar$) terms by $\msbar\equiv \frac{2}{\epsilon}-\gamma+\log(4\pi)$.
 Up to leading order in the coupling constant $\alpha$ we then find the renormalization factor to be
\be\label{Z}
Z_{\phi}^{-1}=1+\left.\frac{\d\Sigma}{\d p^2}\right|_{p^2=-m^2}=1+\frac{\alpha^2}{3(4\pi)^2}\,.
\ee
The total amplitude due to the tree-level scattering is then given by the LSZ reduction formula as
\be\label{tree2}
\mathcal A_{\rm tree}=Z_\phi\times\mathcal A_{\rm tree,0}\,.
\ee

\subsection{One loop}
There are two contributions to the one-loop amplitude of the $\chi\phi\to\chi\phi$ scattering, shown in Figs.~\ref{fig:1b} and \ref{fig:1c}, so that
\be
\mathcal A_{\rm 1-loop}=\mathcal A_{\phi\psi^2}+\mathcal A_{\phi\psi^2h}\,.
\ee

\subsubsection{$\mathcal A_{\phi\psi^2}$ amplitude}\label{sec:phipsi2}
First, let us deal with the loop process supported by the cubic interaction $\phi\psi^2$. This is depicted in Fig.~\ref{fig:1b} and the corresponding amplitude is
\be
\begin{split}
i\mathcal A_{\phi\psi^2}(s,t)=\int \frac{\d^dp}{(2\pi)^d}&V_{\phi\psi^2}^2 \Delta_{\psi}(p)\Delta_\psi(p+k_3)\Delta_{\psi}(p-k_1)\\
&V_M^{\mu\nu}(p+k_3,k_1-p)D_{\mu\nu;\alpha\beta}(k_2+k_4)V_0^{\alpha\beta}(k_2,k_4)\,.
\end{split}
\ee
This gives
\be
\begin{split}
&i\mathcal A_{\phi\psi^2}(s,t)=\frac{16M^2\alpha^2}{\mpl^2t}\int \frac{\d^dp}{(2\pi)^d}\frac{1}{[p^2+M^2]}\frac{1}{[(p+k_3)^2+M^2]}\frac{1}{[(p-k_1)^2+M^2]}\\
&\qquad\times\left[-\frac{s^2}{4}-\frac{st}{4}+\frac{M^2t}{4}+\frac{p^2t}{4}-\frac{1}{2}sp\cdot(k_2-k_4)-\frac{1}{2}tp\cdot(k_1+k_2)+(p\cdot k_2) \,(p\cdot k_4)\right]\,,
\end{split}
\ee
where we have used momentum conservation $k_1+k_2+k_3+k_4=0$. We then combine the three propagators in the denominator by introducing the Feynman parameters $x,y$ and transform the momentum integration variable as $p^\mu\to k^\mu\equiv p^\mu-xk_1^\mu+yk_3^\mu$. This leads to our final expression
\be
i\mathcal A_{\phi\psi^2}(s,t)=\frac{8M^2\alpha^2}{\mpl^2t}\mu^\epsilon\int_0^1\text{d}x\int_0^{1-x}\text{d}y\int\frac{\d^dk}{(2\pi)^d}\frac{A(s,t)+B(t)k^2}{\left[k^2+\Delta\right]^3}\,,
\ee
where we have used $\int\frac{\d^dk}{(2\pi)^d} k^\mu =0$ and $\int\frac{\d^dk}{(2\pi)^d} k^\mu k^\nu =\frac{1}{(2\pi)^d}\frac{1}{d}k^2\eta^{\mu\nu}$ and have defined
\be
\begin{split}
&A(s,t) \equiv -s(s+t)(-1+x+y)^2+M^2t\,,\\
 &B(t)\equiv \frac{(d-2)}{d}t\,,\\
&\Delta\equiv M^2-xyt\,.
\end{split}
\ee
The integral over $k^\mu$ can be taken using the dimensional regularization result \eqref{dimreg}:
\be\label{loop1}
i\mathcal A_{\phi\psi^2}(s,t)=\frac{i8M^2\alpha^2}{\mpl^2t}\int_0^1\text{d}x\int_0^{1-x}\text{d}y\left[\frac{A(s,t)}{32\pi^2\Delta}+\frac{B(t)}{16\pi^2}\left(-\frac{1}{2}+\msbar-\log\frac{\Delta}{\mu^2}\right)\right]\,.
\ee
There are a few important features of this amplitude that need to be discussed. First, it is easy to see that the quantity $\Delta$, appearing in both the denominator and the logarithm in the expression above, vanishes for $t \geq 4M^2$, and implies that there is a branch cut for these values of $t$. While it is thus apparent that the amplitude $\mathcal A_{\phi\psi^2}(s,t)$ is non-analytic for $t\geq 4M^2$ we see that it is polynomial and thus analytic in $s$ (the only $s$-dependence appears in the quantity $A(s,t)$ defined above). Second, as we shall show in detail below the amplitude exhibits a $t$-channel pole, i.e. $\mathcal A_{\phi\psi^2}(s,t)\sim\frac{\alpha^2}{\mpl^2}\frac{s^2}{t}+\dots$. In the low-energy EFT such a contribution to the scattering amplitude can be obtained in the presence of a new operator $\sim \alpha^2\sqrt{-g}(\partial\phi)^2$, corresponding to a redressing of the kinetic term of $\phi$. Indeed, as we shall see, this pole cancels out in the total amplitude for the $\chi\phi\to\chi\phi$ scattering once the wavefunction renormalization of eq.~\eqref{Z} is taken into account. Finally, the amplitude has a finite $s^2$ contribution thus implying that the positivity bounds \eqref{positivity} would give a non-trivial constraint on the parameters of the theory. Let us address the two latter points in detail now.

\begin{itemize}
\item $t$-pole:\\
In order to find the contribution to the $t$-pole, it is sufficient to evaluate the integrands of the full amplitude $\mathcal A_{\phi\psi^2}$ at $t=0$. Since $B(t=0)=0$ this gives
\be\label{tpole1}
i\mathcal A_{\phi\psi^2,\text{ pole}}(s,t)=\frac{i8M^2\alpha^2}{\mpl^2t}\int_0^1\text{d}x\int_0^{1-x}\text{d}y\left.\frac{A(s,t)}{32\pi^2\Delta}\right|_{t=0}\,,
\ee
where $A(s,t=0) \equiv -s^2(-1+x+y)^2$ and $\left.\Delta\right|_{t=0}=M^2$ making it easy to perform the integrals over Feynman variables. As a result we obtain:
\be
\begin{split}
i\mathcal A_{\phi\psi^2,\text{ pole}}(s,t)=-\frac{i\alpha^2}{\mpl^2t}\frac{s^2}{3(4\pi)^2}\,.
\end{split}
\ee
This combines with the result from the tree amplitude in \eqref{tree2} as
\be\begin{split}
i\mathcal A_{t-\text{pole}}&=iZ_\phi\times\mathcal A_{\rm tree,0}+i\mathcal A_{\phi\psi^2,\text{ pole}}=-\frac{is^2}{4\mpl^2t}+\mathcal O(t^0)\,,
\end{split}
\ee
and we see that the $\mathcal O(\alpha^2)$ contribution to the $t$-pole has cancelled leaving only the original tree-level pole. From the definition of $A(s,t)$ and $B(t)$ we see that all their next order contributions are proportional to $t$ and the amplitude is thus finite in the forward limit.


\item $s^2$ contribution:\\
To find the relevant $s^2$ contribution to the total amplitude \eqref{loop1} it is sufficient to expand the integrand around $t=0$. Indeed, we know that the amplitude is analytic for $t\leq 4M^2$ and that the positivity bounds will be applied in the forward limit with $t=0$. This makes the small $t$ expansion a valid approximation. We are then only interested in the $s^2t$ contribution in the integrand which cancels the $t$ in the overall denominator and leads to a finite $s^2$ contribution to the scattering amplitude. Since the only $s^2$ dependence is in the quantity $A(s,t)$ we obtain:
\be\label{loops2}
\begin{split}
i\mathcal A_{\phi\psi^2,\,s^2}(s,t)&=\frac{i8M^2\alpha^2}{\mpl^2t}\int_0^1\text{d}x\int_0^{1-x}\text{d}y\left.\frac{-A(s,t)}{32\pi^2\Delta^2}\right|_{t=0}\times\frac{\d\Delta}{\d t}\times t\\
&=-\frac{i\alpha^2s^2}{90(4\pi)^2M^2\mpl^2}+\dots\,,
\end{split}
\ee
where the ellipsis stands for terms that do not have any $s^2$ contribution.

\end{itemize}

\subsubsection{$\mathcal A_{\phi\psi^2h}$ amplitude}
The last step in computing the total scattering amplitude for the process $\chi\phi\to\chi\phi$ is determining the contribution  from the one-loop process shown in Fig.~\ref{fig:1c}. This is given by
\be
\begin{split}
i\mathcal A_{\phi\psi^2h}(s,t)&=2\times\frac{1}{2}\int\frac{\d^dp}{(2\pi)^d}V_{\phi\psi^2}\Delta_{\psi}(p)\Delta_\psi(p+k_3)V_{\phi\psi^2h}^{\mu\nu}D_{\mu\nu;\alpha\beta}(k_2+k_4)V_0^{\alpha\beta}(k_2,k_4)\\
&=-\frac{4\alpha^2M^2}{\mpl^2}\mu^\epsilon\int\frac{\d^dp}{(2\pi)^d}\frac{1}{[p^2+M^2]}\frac{1}{[(p+k_3)^2+M^2]}\,,
\end{split}
\ee
where again the $1/2$ is the symmetry factor of the diagram and there is an extra factor of $2$ as the loop can be on either external $\phi$ leg.
After introducing the Feynman parameter $x$, shifting the momentum to $k^\mu\equiv p^\mu+xk_3^\mu$ and taking the integral as in dimensional regulation in \eqref{dimreg} for $d=4-\epsilon$ we obtain
\be
i\mathcal A_{\phi\psi^2h}(s,t)=-\frac{4i\alpha^2M^2}{(4\pi)^2\mpl^2}\int_0^1\text{d}x\left[\msbar-\log\frac{M^2}{\mu^2}\right]\,.
\ee
This amplitude is independent on both $s$ and $t$ and thus has no impact on the positivity bounds.

\section{Compactified Amplitudes in the Partial UV Theory}\label{sec:3d}
\begin{figure}[t]
    \centering
\includegraphics[height=1.5in]{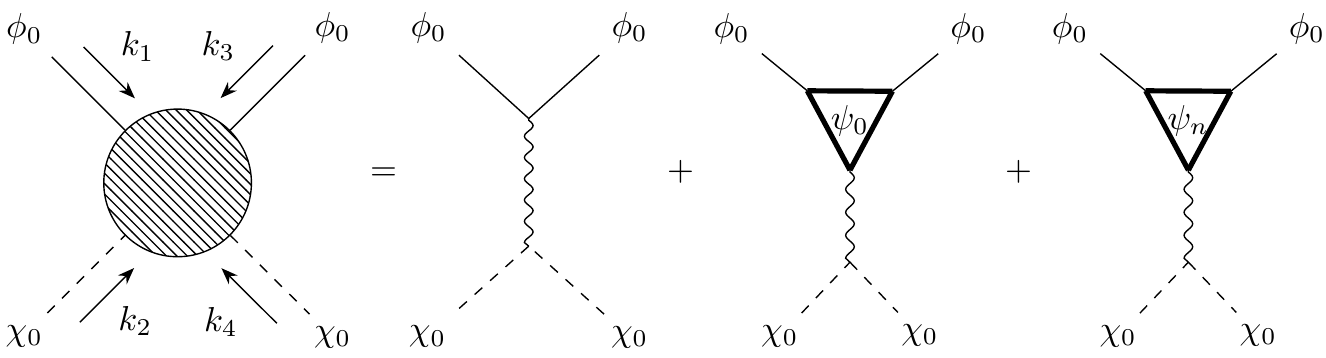}
\caption{Principal contributions to $\chi_0\phi_0\to\chi_0\phi_0$ scattering. Here $n\in [1,\infty)$. }
\label{fig:scattering3}
\end{figure}

\begin{figure}[t]
    \centering
\includegraphics[height=1.5in]{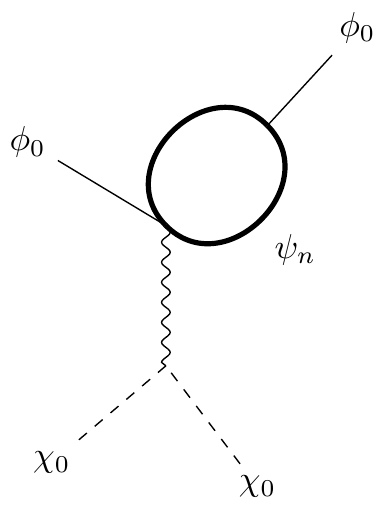}
\caption{Additional contributions to $\chi_0\phi_0\to\chi_0\phi_0$ scattering. Here $n\in [0,\infty)$. }
\label{fig:KKdiagram}
\end{figure}

We now apply explicitly the procedure suggested in  Ref.~\cite{Bellazzini:2019xts} to our model and show how the contradiction~\eqref{contradiction} manifests itself in that formalism. We compactify one of the spatial directions (denoted here by $z$)  on a circle of length $L$. In this section we denote the 4d metric by $\hat g_{MN}$ where $N,M = 0, \dots, 3$ and parameterize it as
\be
\hat g_{MN}=e^{-\sigma}\begin{pmatrix}
e^{2\sigma}g_{\mu\nu}+V_\mu V_\nu\ \ &\ \ V_\mu\\[5pt]
V_\nu\ \ &\ \ 1
\end{pmatrix}\,,
\ee
where the Greek indices $\mu,\nu$ are the 3d ones. The heavy scalar $\psi$ can be expanded into KK modes
\be
\psi(x^M)=\sum_n e^{\frac{2\pi inz}{L}}\psi_n(x^\mu)\,,
\ee
so that $\psi(z+L)=\psi(z)$. A similar decomposition can be used for all the other fields we are interested in.
However, in what follows we shall focus on scatterings of the zero KK modes $\phi_0 \chi_0 \to \phi_0 \chi_0$ for which (to the order we are working in), only the zero modes of the gravitational fields $\sigma, V_{\mu}, g_{\mu\nu}$ contribute. Without loss of generality and in order to avoid unnecessarily heavy notation, in what follows we shall therefore simply denote by  $\sigma, V_{\mu}, g_{\mu\nu}$ the zero modes of the gravitational fields. Focusing on the zero KK modes for all fields aside from the heavy scalar $\psi$ and integrating our model~\eqref{pUV_0} over the compactified direction we thus get
\be
\begin{split}
S_{\rm pUV, 3d}&=L\int \d^3x\sqrt{-g}\Bigg\{\frac{\mpl^2}{2}\left(R-\frac{1}{2}(\partial\sigma)^2-\frac{1}{4}V_{\mu\nu}V^{\mu\nu}e^{-2\sigma}\right)-\frac{1}{2}(\partial\chi_0)^2\\
&-\frac{1}{2}(\partial\phi_0)^2-\frac{1}{2}(\partial\psi_0)^2-\frac{1}{2}M^2e^\sigma\psi_0^2-\alpha M e^\sigma\phi_0\psi_0^2-2\alpha M\sum_{n=1}^\infty e^\sigma \phi_0\psi_n\psi_{n}^\dagger\\
&-\sum_{n=1}^\infty \left[g^{\mu\nu}\partial_\mu\psi_n\partial_\nu\psi_{n}^\dagger+M^2e^\sigma\psi_n\psi^\dagger_n-2\frac{in}{L}V^\mu\psi_n\partial_\mu\psi_{n}^\dagger+\frac{n^2}{L^2}(e^{2\sigma}+V_\alpha V^\alpha)\psi_n\psi_{n}^\dagger\right]\Bigg\}\,,
\end{split}
\ee
where now $(\partial\chi_0)^2\equiv g^{\mu\nu}\partial_\mu\chi_0\partial_\nu\chi_0$ etc. with $\mu,\nu = 0,1,2$ and $g_{\mu\nu}$ is the 3d metric. We have also defined $V_{\mu\nu}\equiv\partial_\mu V_\nu-\partial_\nu V_\mu$ and are treating the KK modes $\psi_n$ as complex by identifying $\psi_{-n}=\psi^\dagger_n$\,.
 For the one-loop $\chi_0\phi_0\to\chi_0\phi_0$ scattering the relevant terms in the above action are only
\ba
S_{\rm UV, 3d}&\supset L&\int \d^3x\sqrt{-g}\Bigg\{\frac{\mpl^2}{2}R-\frac{1}{2}(\partial\chi_0)^2-\frac{1}{2}(\partial\phi_0)^2-\frac{1}{2}(\partial\psi_0)^2-\frac{1}{2}M^2\psi_0^2\\
&-&\alpha M \phi_0\psi_0^2-2\alpha M\sum_{n=1}^\infty  \phi_0\psi_n\psi_{n}^\dagger-\sum_{n=1}^\infty \left(g^{\mu\nu}\partial_\mu\psi_n\partial_\nu\psi_{n}^\dagger+\left(M^2+\frac{4\pi^2n^2}{L^2}\right)\psi_n\psi^\dagger_n\right)\Bigg\}\,.\nn
\ea
The scalar fields $\psi_n$ acquire an effective mass $M_n$ defined as
\be\label{massn}
M_n^2\equiv M^2+\frac{4\pi^2n^2}{L^2}\,.
\ee
The diagrams contributing to the  $\chi_0\phi_0\to\chi_0\phi_0$ scattering up to one-loop order are shown in Fig.~\ref{fig:scattering3}. Since we are only interested in contributions to the scattering amplitude that grow as $s^2$ we have dropped other diagrams, like the one in Fig.~\ref{fig:KKdiagram}. The computation of the scattering amplitude is very similar to the 4d case shown in great detail in appendix~\ref{sec:uv_ampl}.  Here we only show the main results of the computation. Taking care of the appropriate canonical normalizations the vertices expressed in \eqref{rules} get  rescaled by  $V_{\phi_0\psi_n^2}=-2i\alpha M/\sqrt{2\pi L}$ and $V_{M_n}^{\mu\nu}=V_{M_n}^{\mu\nu}/\sqrt{2\pi L}$.
Using these rules in 3d, the tree-level amplitude gives
\be
i\mathcal A_{\rm tree,0}(s,t)=-\frac{i}{\mpl^2L}\frac{s^2+st}{t}\,,
\ee
and up to the factor $1/L$ coincides with the 4d result given in \eqref{amp_pUV}. Most importantly, it also exhibits a $t$-pole. Taking into account the shift in the self-energy of $\phi_0$ due to the $\psi_0$ and $\psi_n$ loops the tree-level amplitude receives $\alpha^2$ corrections as $\mathcal A_{\rm tree}=Z_{\phi_0}\times\mathcal A_{\rm tree,0}$ where $Z_{\phi_0}$ is the wavefunction renormalization factor
\be
Z_{\phi_0}=1-\frac{\alpha^2M^2}{48\pi L}\sum_{n=-\infty}^\infty\frac{1}{M_n^3}+\mathcal O(\alpha^4)\,.
\ee
The amplitude for the loop processes in Fig.~\ref{fig:scattering3} can be written for both $n=0$ and $n\neq 0$ as
\be
\begin{split}
i\mathcal A_{\phi_0\psi_n^2}(s,t)=\mathcal N\int \frac{\d^3p}{(2\pi)^3}&V_{\phi_0\psi_n^2}^2 \Delta_{\psi_n}(p)\Delta_{\psi_n}(p+k_3)\Delta_{\psi_n}(p-k_1)\\
&V_{M_n}^{\mu\nu}(p+k_3,k_1-p)D_{\mu\nu;\alpha\beta}(k_2+k_4)V_0^{\alpha\beta}(k_2,k_4)\,,
\end{split}
\ee
where the symmetry factor $\mathcal N = 1$ for $n=0$ and $\mathcal N=2$ for $n\ge 1$. Using dimensional-regularization,
this gives
\be\label{psinloop}
i\mathcal A_{\phi_0\psi_n^2}(s,t)=-\frac{8\mathcal N\alpha^2M^2}{L^2\mpl^2t}\int_0^1 \d x\int_0^{1-x} \text {d}y\int\frac{\d^3k}{(2\pi)^3}\,\frac{(s^2+st)(-1+x+y)^2-\frac{(d-2)}{d}k^2t-M_n^2t}{[k^2+M_n^2-xyt]^3}\,,
\ee
where $x,y$ are the Feynman parameters and we have shifted the momentum as $k^\mu\equiv p^\mu-xk_1^\mu+yk_3^\mu$. After performing the integration over momenta, as in subsection \ref{sec:phipsi2} we find that there is an $s^2/t$ pole:
\be
\mathcal A_{\phi_0\psi_n^2,\text{ pole}}(s,t)=-\frac{\alpha^2M^2}{48\pi\mpl^2 L^2}\sum_{n=-\infty}^\infty\frac{s^2}{t}\frac{1}{M_n^3}\,,
\ee
that cancels out when added to the tree-level contribution leaving, as in 4d,
\be
\mathcal A_{t-\text{pole}}=Z_{\phi_0}\times\mathcal A_{\rm tree,0}+\mathcal A_{\phi\psi_n^2,\text{ pole}}=-\frac{(s^2+st)}{\mpl^2Lt}+\mathcal O(t^0)\,.
\ee
We then find the $s^2$ contribution --- relevant for the positivity bounds \eqref{positivity} --- by evaluating the integrals in \eqref{psinloop} in $t\to 0$ limit. Combined together with the tree-level result we obtain the final expression of the regularized $\chi_0\phi_0\to\chi_0\phi_0$ scattering amplitude:
\be\label{amp_pUV2}
\mathcal A_{\rm pUV, 3d}(s,t)=-\frac{s^2}{\mpl^2Lt}-\frac{\alpha^2M^2}{240(4\pi)L^2\mpl^2}\sum_{n=-\infty}^\infty s^2\frac{1}{M_n^5}+\mathcal O(t^0)\,.
\ee
As a consistency check, we can take the continuum limit of the above amplitude by sending $L\to\infty$ in the expression $\mathcal A_{\rm pUV, 3d}(s,t)\times L$. Using the relation $\lim_{L \rightarrow \infty}\sum_{n=-\infty}^\infty M_n^{-5}=\frac{2L}{3\pi M^4}$ we indeed recover~\eqref{amp_pUV}.\\

We note that formally the pole $s^2/t$ is also present in the regularized amplitude \eqref{amp_pUV2}. However, since there are no propagating massless spin-2 fields in 3d, this pole cannot be physical. This can be seen in the eikonal approximation as discussed in section~\ref{Infrared}. Roughly speaking the eikonal resummation of the $t$-channel pole in \eqref{amp_pUV2} amounts to the replacement \eqref{eikonal104}
\be
\frac{s u}{\mpl^2Lt} \rightarrow  \frac{su}{\mpl^2L(t -4 s u/(\mpl^4 L^2)) } \, ,
\ee
which remains finite as $t \rightarrow 0$ and furthermore asymptotes to a constant at large $s$ for fixed $t$. For this approximation to work, the forward limit must be taken before decompactifying the $z$-direction by $L\to\infty$. In other words, one must ensure that $t\ll 1/L^2$. Following the prescription of  \cite{Bellazzini:2019xts} one can then further argue that the eikonal approximation will bear no effect on the contributions to the scattering amplitude from the second term in \eqref{amp_pUV2}. This can be  easily understood from the partial wave expansion of the finite terms in the amplitude \eqref{amp_pUV2}. When expanded in terms of partial waves, these only have contributions at low $\ell$. In contrast, the eikonal approximation is dominated by the resummation of the large $\ell$ partial waves of the total amplitude and thus leaves specific small $\ell$ contributions unscathed. The validity of applying positivity bounds in 3d is critically discussed in section~\ref{Infrared}, however assuming for now that they would be applicable, we would infer
\be\label{contradiction2}
\frac{\text{d}^2\mathcal A_{\rm pUV, 3d}(s,0)}{\text{d}s^2}=-\frac{\alpha^2}{120(4\pi)M\mpl^2}\times \frac{1}{L^2M^2}\left(1+2\sum_{n=1}^\infty\frac{1}{\left(1+\frac{4\pi^2n^2}{L^2M^2}\right)^{5/2}}\right)>0\,.
\ee
Clearly this cannot be satisfied for any choice of partial UV completion parameters.

\section{Adding interactions to the UV theory}\label{sec:addUV}
In this section we shall contemplate the possibility that our partial UV completion is to be blamed and establish whether adding other operators that would otherwise be renormalizable in the absence of gravity could help satisfying the compactified positivity bounds. Following this approach, we consider cubic and quartic non-derivative operators that introduce additional $\phi,\psi$ interactions in our UV theory \eqref{pUV_0} and discuss their implications for the positivity bounds applied to the $\chi\phi\to\chi\phi$ scattering amplitude.

\subsection{Cubic operators}
\begin{figure}[t]
\centering
\begin{subfigure}{0.3\textwidth}
\centering
\includegraphics[height=1.5in]{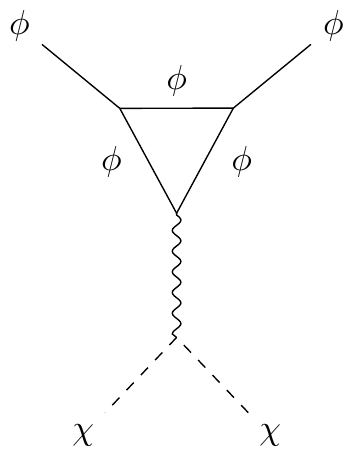}
\caption{One-loop diagram for $\mathcal A_{\phi^3,1}$.} \label{fig:4a}
\end{subfigure}
\hspace*{0.8in} 
\begin{subfigure}{0.3\textwidth}
\centering
\includegraphics[height=1.5in]{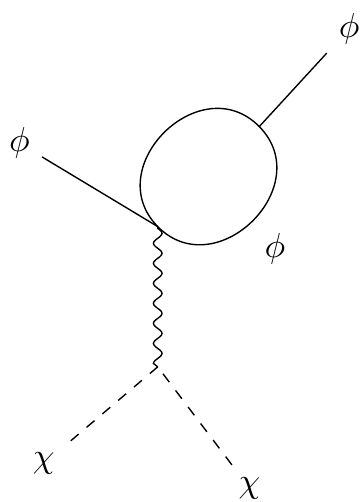}
\caption{One-loop diagram for $\mathcal A_{\phi^3,2}$.} \label{fig:4d}
\end{subfigure}
\caption{Feynman diagrams contributing to scattering processes $\chi\phi\to\chi\phi$ from the additional $-aM\frac{\phi^3}{3!}$ operator in \eqref{cubic}, up to one loop. The wiggly lines correspond to graviton propagators.} \label{fig:4v2}
\end{figure}

\begin{figure}
\centering
\begin{subfigure}{0.3\textwidth}
\centering
\includegraphics[height=1.5in]{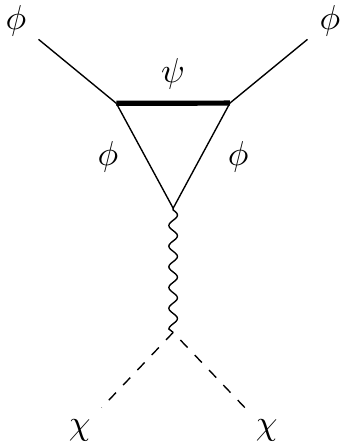}
\caption{One-loop diagram for $\mathcal A_{\psi\phi^2,1}$.} \label{fig:4b}
\end{subfigure}
\hspace*{0.8in} 
\begin{subfigure}{0.3\textwidth}
\centering
\includegraphics[height=1.5in]{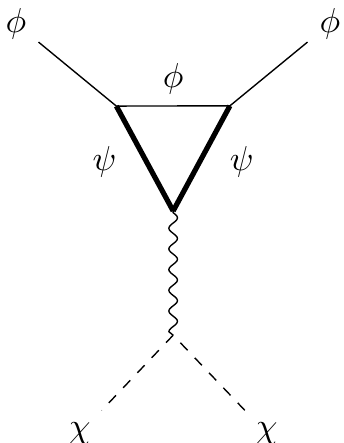}
\caption{One-loop diagram for $\mathcal A_{\psi\phi^2,2}$.} \label{fig:4c}
\end{subfigure}
\caption{The Feynman diagrams for the one-loop scattering processes contributing to $\chi\phi\to\chi\phi$ from the additional cubic EFT operators in \eqref{cubic}. Wiggly lines correspond to graviton propagators, while bold lines correspond to the propagators of the heavy field $\psi$.} \label{fig:4}
\end{figure}

We start by supplementing our model with the following additional cubic interactions
\be\label{cubic}
\mathcal L_{(3)}=\sqrt{-g}\left[-aM\frac{\phi^3}{3!} - bM\psi\frac{\phi^2}{2}\right]\,,
\ee
where $a,b$ are dimensionless constants and we have fixed the mass scale in front of the operators to coincide with the mass of the heavy scalar, $M$. The new processes contributing to the $\chi\phi\to\chi\phi$ scattering are shown in Figs.~\ref{fig:4v2} and \ref{fig:4} and we define the corresponding interaction vertices as $V_{\phi^3}=-iaM$ and $V_{\psi\phi^2}=-ibM$.

\subsubsection{The $\phi^3$ interaction}\label{sec:phi3}
The presence of the $\phi^3$ interaction allows for two more one-loop scattering channels for the $\chi\phi\to\chi\phi$ scattering shown in Figs.~\eqref{fig:4a} and \eqref{fig:4d}. Together with the tree-level amplitude, the total scattering amplitude due to the $\phi^3$ vertex is given by the sum:
\be
\mathcal A_{\phi_3}=Z_\phi\times\mathcal A_{\rm tree,0}+\mathcal A_{\phi_3,1}+\mathcal A_{\phi_3,2}\,,
\ee
where $Z_\phi$ is the corresponding wavefunction normalization factor. Due to the presence of the light loops we add the mass term $-\frac{1}{2}m^2\phi^2$ to the total action. Let us emphasize that, when calculating the amplitudes, the mass appearing in the propagators and in the vertices is this bare mass $m$. On the other hand, when substituting the external momenta we use $k_1^2=k_3^2=-m_{\rm ph}^2$ and $u=-s-t+2m_{\rm ph}^2$ where $m_{\rm ph}$ is the physical mass. We determine the relation between the two below.

\paragraph{Renormalization.} The $\phi^3$ interaction renormalizes the mass and the kinetic term of $\phi$ while the cubic coupling $aM$ does not get renormalized in 4d. Let us find the renormalized quantities up to one-loop order. For this we first find the self-energy up to quadratic order in the coupling $aM$:
\be
\begin{split}
\Sigma(p^2)=\vcenter{\hbox{\includegraphics[width=2.5cm]{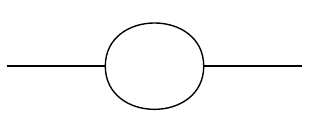}}}=-\frac{a^2M^2}{32\pi^2}\int_0^1\text{d}x\,\left[\msbar+\log\frac{\mu^2}{m^2+x(1-x)p^2}\right]+\mathcal O(a^4M^4)\,.
\end{split}
\ee
Under the assumption $-p^2\leq 4m^2$ this can be integrated to
\begin{align}
&\Sigma(p^2)=-\frac{a^2M^2}{32\pi^2}\left[\log\frac{\mu^2}{m^2}+f(p^2)\right]-\frac{a^2M^2}{32\pi^2}\msbar +\mathcal O(a^4M^4)\,,\\\label{fp}
&f(p^2)\equiv 2-2\sqrt{-\frac{4m^2}{p^2}-1}\arctan\frac{1}{\sqrt{-\frac{4m^2}{p^2}-1}}\,.
\end{align}
We determine the physical mass from the relation
\be\label{phys_mass2}
\left[p^2+m^2+\Sigma(p^2)\right]_{p^2=-m_{\text{ph}}^2}=0\,.
\ee
Substituting the above expression for $\Sigma(p^2)$ we find the exact expression for the one-loop mass renormalization:
\be\label{mphys}
m_{\rm ph}^2=m^2+\frac{a^2M^2}{32\pi^2}\left(\log\frac{m^2}{\mu^2}-f(-m_{\rm ph}^2)-\msbar\right)+\mathcal O(a^4M^4)\,.
\ee
Since in fact $m^2=m_{\rm ph}^2+\mathcal O(a^2M^2)$ we can substitute $m^2=m^2_{\rm ph}$ in the logarithm above and use that also $f(-m_{\rm ph}^2)=f(-m^2)+\mathcal O(a^2M^2)=2-\frac{\pi}{\sqrt{3}}$ and derive the one-loop renormalization group (RG) equation for the bare mass:
\be
\frac{\text{d}m^2}{\text{d}\log\mu}=\frac{a^2M^2}{16\pi^2}+\mathcal O(a^4M^4)\,,
\ee
needed to ensure that the physical mass does not depend on the renormalization scale. The latter equation can be solved for $m$ to find $m^2=\frac{a^2M^2}{16\pi^2}\log(c\mu)$ where $c$ is an integration constant. Substituting this in \eqref{mphys} we obtain
\be
m_{\rm ph}^2=\frac{a^2M^2}{32\pi^2}\log\frac{m^2}{\tilde c^2}\,,
\ee
where we have redefined the integration constant as $\log \tilde c=-\log c+2-\frac{\pi}{\sqrt{3}}+\msbar$.

Similarly, we find the wavefunction renormalization factor
\be
Z_{\phi}^{-1}=1+\left.\frac{\d\Sigma}{\d p^2}\right|_{p^2=-m^2}=1-\frac{a^2M^2}{32\pi^2}\frac{1}{m_{\rm ph}^2}\left[1-\frac{4m^2}{m_{\rm ph}^2}\frac{1}{\sqrt{\frac{m^2}{m_{\rm ph}^2}-\frac{1}{4}}}\arctan\frac{1}{\sqrt{\frac{m^2}{m_{\rm ph}^2}-\frac{1}{4}}}\right]+\mathcal O(a^4M^4)\,.
\ee
Using again that $m_{\rm ph}^2=m^2+\mathcal O(a^2M^2)$ we obtain up to one-loop order
\be
Z_\phi=1-\frac{a^2M^2}{32\pi^2}\left(\frac{-9+2\sqrt{3}\pi}{9m_{\rm ph}^2}\right)\,.
\ee

\paragraph{The tree-level amplitude.}
The tree-level amplitude of the $\chi\phi\to\chi\phi$ scattering is given by $\mathcal A_{\rm tree}=Z_\phi\times\mathcal A_{\rm tree,0}$ with
\be
\mathcal A_{\rm tree, 0}=-\frac{1}{\mpl^2}\frac{(s-m_{\rm ph}^2)^2-(s-m^2)t}{t}\,,
\ee
where we have made the distinction between the physical and bare masses entering the above relation. Expressing the bare mass in terms of the physical mass as in \eqref{mphys} we obtain the following expression for the full tree-level amplitude with $\mathcal O(a^2M^2)$ corrections
\be\label{amp0}
\begin{split}
\mathcal A_{\rm tree}=\frac{(m_{\rm ph}^2-s)(m_{\rm ph}^2-u)}{\mpl^2t}&\left(1-\frac{a^2M^2}{32\pi^2}\left(\frac{-9+2\sqrt{3}\pi}{9m_{\rm ph}^2}\right)\right)\\
-\frac{a^2M^2}{32\pi^2\mpl^2}&\left(\log\frac{m_{\rm ph}^2}{\mu^2}-f(-m_{\rm ph}^2)-\msbar\right)\,.
\end{split}
\ee

\paragraph{The first loop process.} The scattering amplitude for the process in Fig.~\ref{fig:4a} schematically reads
\be
\begin{split}
  i\mathcal A_{\phi^3,1}(s,t)=\int\frac{\text{d}^{d}p}{(2\pi)^{d}}&V_{\phi^3}^2\Delta_{\phi}(p)\Delta_{\phi}(k_{1}-p)\Delta_{\phi}(p+k_{3})\\
  &V_m^{\mu\nu}(p+k_3,k_1-p)D_{\mu\nu;\alpha\beta}(k_{2}+k_{4})V_0^{\alpha\beta}(k_2,k_4)\,.
  \end{split}
\ee
Performing the simplifications needed and transforming the integration momentum as $k^\mu\equiv p^\mu-(x k_1^\mu-yk_3^\mu)$ leads to
\be\label{integrand}
\begin{split}
i\mathcal A_{\phi^3,1}(s,t)=\frac{2M^2a^2}{\mpl^2t}\mu^\epsilon\int_0^1\text{d}x\int_0^{1-x}\text{d}y\int\frac{\text{d}^dk}{(2\pi)^d}\frac{A(s,t)+B(t)k^2}{\left[k^2+\Delta(t)\right]^3}\,,
\end{split}
\ee
where we have again defined:
\be
\begin{split}
&A(s,t) \equiv -\left[(s-m_{\rm ph}^2)^2+st\right](-1+x+y)^2+m^2t\,,\\
 &B(t)\equiv \frac{(d-2)}{d}t\,,\\
&\Delta(t)\equiv m^2+m_{\rm ph}^2(-1+x+y)(x+y)-xyt\,.
\end{split}
\ee
As before, we shall use $m_{\rm ph}^2=m^2+\mathcal O(a^2M^2)$. On evaluating the momentum integral we have
\be\label{mneq0}
\begin{split}
\lim_{\epsilon\to0}\mathcal A_{\phi^3,1}=\frac{2a^2M^2}{\mpl^2t}\int_0^1\text{d}x\int_0^{1-x}\text{d}y\left[\frac{A(s,t)}{32\pi^2\Delta}+\frac{t}{32\pi^2}\left(-1+\msbar+\log\frac{\mu^2}{\Delta}\right)\right]\,,
\end{split}
\ee
where we have also used the explicit form of $B(t)$ since it involved an additional $d$-dependence. We thus obtain for the full one-loop amplitude for the process in Fig.~\ref{fig:4a}:
\be\label{amp1}
\begin{split}
\mathcal A_{\phi^3,1}(s,t)=&-\frac{a^2M^2}{16\pi^2\mpl^2}\frac{\left((s-m_{\rm ph}^2)^2+st\right)}{t}\int_0^1\text{d}x\int_0^{1-x}\text{d}y\,\frac{(-1+x+y)^2}{\Delta(t)}\\
&+\frac{a^2M^2}{32\pi^2\mpl^2}\left[-1+\msbar+\log\frac{\mu^2}{m_{\rm ph}^2}+2\int_0^1\text{d}x\int_0^{1-x}\text{d}y\,\left(\frac{m_{\rm ph}^2}{\Delta(t)}+\log\frac{m_{\rm ph}^2}{\Delta(t)}\right)\right]\,.
\end{split}
\ee
The analysis of the amplitude in \eqref{mneq0} then continues in a manner very similar to section~\ref{sec:phipsi2} where the cubic $\phi\psi^2$ interaction was analyzed in detail. As before we note that the amplitude has a branch cut in the complex $t$-plane starting from the point where the denominator $\Delta$ becomes negative. In the expression for $\Delta$ given above it is apparent that in the integration region the first two terms are greater than zero and also $xy\geq0$ therefore the branch cut lies along the real positive values of $t$. In particular, we find that there is a branch cut for $t\geq 4m_{\rm ph}^2$. As before when analyzing the $\phi\psi^2$ interaction, we find that the first line of the one-loop amplitude $\mathcal A_{\phi^3,1}$ has both a $t$-channel pole and an $s^2$ contribution. The second line is independent on $s,t$, but depends on the renormalization scale. There is no $s$-dependence on the third line of the amplitude, moreover, it is finite and non-zero in $t\to 0$ limit. Hence, it can be easily disregarded in the context of positivity bounds.

\paragraph{The second loop amplitude.} The scattering amplitude for the process in Fig.~\ref{fig:4d} reads
\be
\begin{split}
i\mathcal A_{\phi^3,2}(s,t)&=\int\frac{\d^dp}{(2\pi)^d}V_{\phi^3}\Delta_{\phi}(p)\Delta_\phi(p+k_3)V_{\phi^3h}^{\mu\nu}D_{\mu\nu;\alpha\beta}(k_2+k_4)V_0^{\alpha\beta}(k_2,k_4)\\
&=-\frac{a^2M^2}{\mpl^2}\mu^\epsilon\int_0^1\text{d}x\int\frac{\d^dk}{(2\pi)^d}\frac{1}{[k^2+m^2-m_{\rm ph}^2x(1-x)]^2}\,,
\end{split}
\ee
where $k^\mu\equiv p^\mu+xk_3^\mu$ and the vertex $V_{\phi^3h}^{\mu\nu}\equiv \frac{-iaM}{2\mpl}\eta^{\mu\nu}$. Integrating we obtain
\be\label{amp2}
\mathcal A_{\phi^3,2}(s,t)=-\frac{a^2M^2}{16\pi^2\mpl^2}\left[\msbar+\log\frac{\mu^2}{m_{\rm ph}^2}+f(-m_{\rm ph}^2)\right]\,,
\ee
where $f(-p^2)$ is defined in \eqref{fp}.

\paragraph{The total amplitude.} Adding all the contributions in \eqref{amp0}, \eqref{amp1}, \eqref{amp2} we get the total amplitude for the $\chi\phi\to\chi\phi$ scattering in $\phi^3$ theory. As expected, the prefactors to the scale-dependent terms cancel out,  $\frac{a^2M^2}{16\pi^2\mpl^2}\log\frac{\mu^2}{m_{\rm ph}^2}\left(\frac{1}{2}+\frac{1}{2}-1\right)=0$ and so do the $\msbar$ terms. The total amplitude thus becomes
\ba
\mathcal A_{\phi^3}(s,t)&=&-\frac{(m_{\rm ph}^2-s)(m_{\rm ph}^2-s-t)}{\mpl^2t}\\
&+&\frac{a^2M^2}{32\pi^2}\frac{\left((s-m_{\rm ph}^2)^2+st\right)}{\mpl^2t}\left[\frac{-9+2\sqrt{3}\pi}{9m_{\rm ph}^2}-2\int_0^1\text{d}x\int_0^{1-x}\text{d}y\,\frac{(-1+x+y)^2}{\Delta(t)}\right]\nn \\
&-&\frac{a^2M^2}{32\pi^2\mpl^2}\left[f(-m_{\rm ph}^2)+\frac{2\sqrt{3}\pi}{9}-2\int_0^1\text{d}x\int_0^{1-x}\text{d}y\,\left(\frac{m_{\rm ph}^2}{\Delta(t)}+\log\frac{m_{\rm ph}^2}{\Delta(t)}\right)\right]+\mathcal O(a^4M^4)\,,\nn
\ea
where as before $f(-m_{\rm ph}^2)=2-\frac{\pi}{\sqrt{3}}+\mathcal O(a^4M^4)$. The first line contains the pure tree-level result. All the $s$-dependence and the apparent one-loop contribution to the $t$-channel pole appear on the second line, while the third line only depends on $t$. We see from here that the one-loop $t$-channel pole present on the second line cancels out exactly: when the integrand on the second line is evaluated at $t=0$ the whole contribution in the square brackets equals to zero. Similarly, the term on the third line vanishes once the integral is evaluated at $t=0$. For the forward limit answer we thus have
\be\label{amp_phi3}
\begin{split}
\lim_{t\to 0}\mathcal A_{\phi^3}(s,t)=&-\frac{(m_{\rm ph}^2-s)^2}{\mpl^2t}+\frac{(m_{\rm ph}^2-s)}{\mpl^2}+\frac{a^2M^2}{32\pi^2}\frac{(s-m_{\rm ph}^2)^2}{\mpl^2}\frac{(-45+8\sqrt{3} \pi)}{162m_{\rm ph}^4}+\mathcal O(a^4M^4)\,,
\end{split}
\ee
where $-45+8\sqrt{3}\pi=-1.49$. As for the one-loop amplitude due to the $\phi\psi^2$ interaction for the process shown in Fig.~\ref{fig:1b} the contribution to the $s^2$ term is negative.

\subsubsection{The $\psi\phi^2$ interaction}
The addition of the $\psi\phi^2$ interaction leads to two new diagrams contributing to the $\chi\phi\to\chi\phi$ scattering, shown in  Figs.~\ref{fig:4b} and \ref{fig:4c}. The analytic properties of the corresponding scattering amplitudes are very similar to what was discussed for the QED interaction term $-\alpha M\phi\psi^2$ discussed in section~\ref{sec:phipsi2}. We briefly analyze the amplitudes for the new processes below.

\begin{itemize}
\item The process in Fig.~\ref{fig:4b}.
For this scattering process the amplitude reads
\be
i\mathcal A_{\psi\phi^2,1}(s,t)=\int\frac{\text{d}^{d}p}{(2\pi)^{d}}V_{\psi\phi^2}^2\Delta_{\psi}(p)\Delta_{\phi}(k_{1}-p)\Delta_{\phi}(p+k_{3})V_m^{\mu\nu}D_{\mu\nu;\alpha\beta}(k_{2}+k_{4})V_0^{\alpha\beta}\,.
\ee
It can again be manipulated in the familiar form
\be
\begin{split}
i\mathcal A_{\psi\phi^2,1}(s,t)&=\frac{2M^2b^2}{\mpl^2t}\mu^\epsilon\int_0^1\text{d}x\int_0^{1-x}\text{d}y\int\frac{\text{d}^dk}{(2\pi)^d}\frac{A(s,t)+B(t)k^2}{\left[k^2+\Delta\right]^3}\,,
\end{split}
\ee
and we have defined
\be
\begin{split}
&A(s,t) \equiv -s(s+t)(-1+x+y)^2\,,\\
 &B(t)\equiv \frac{(d-2)}{d}t\,,\\
&\Delta\equiv -M^2(-1+x+y)-xyt\,.
\end{split}
\ee
Again this amplitude carries  both a $s^2/t$ pole and a regular $s^2$ contribution. Expanding the integrand abound $t=0$ we obtain
\be
\mathcal A_{\psi\phi^2,1}(s,t)=-\frac{b^2}{(4\pi)^2M^2\mpl^2}\int_0^1\text{d}x\int_0^{1-x}\text{d}y\,xy\times s^2+\mathcal O(s)\,,
\ee
where $\mathcal O(s)$ stands for all the other terms in the amplitude, growing with at most one power of~$s$. The integral over the Feynman parameters gives $1/24$. Hence we conclude that the $s^2$ contribution to the scattering amplitude $\mathcal A_{\psi\phi^2,1}$ is again negative.

\item The process in Fig.~\ref{fig:4c}.
Similarly, for the second diagram, the amplitude reads again
\be
\begin{split}
i\mathcal A_{\psi\phi^2,2}(s,t)&=\frac{2M^2b^2}{\mpl^2t}\mu^\epsilon\int_0^1\text{d}x\int_0^{1-x}\text{d}y\int\frac{\text{d}^dk}{(2\pi)^d}\frac{\tilde A(s,t)+\tilde B(t)k^2}{\left[k^2+\tilde \Delta\right]^3}\,,
\end{split}
\ee
where this time
\be
\begin{split}
&\tilde A(s,t) \equiv -s(s+t)(-1+x+y)^2+M^2t\,,\\
 &\tilde B(t)\equiv \frac{(d-2)}{d}t\,,\\
&\tilde\Delta\equiv M^2(x+y)-xyt\,.
\end{split}
\ee
Again, expanding the integrand abound $t=0$ we obtain  the following $s^2$ contribution to the scattering amplitude
\be
\mathcal A_{\psi\phi^2,2}(s,t)=-\frac{b^2}{(4\pi)^2M^2\mpl^2}\int_0^1\text{d}x\int_0^{1-x}\text{d}y\,\frac{xy(-1+x+y)^2}{(x+y)^2}\times s^2+\mathcal O(s)\,.
\ee
Also here the integral is positive and for $m=0$ integrates to $1/72$, leading to a negative contribution to the amplitude.

\end{itemize}

As for the $s^2/t$ pole appearing in both amplitudes calculated above, it vanishes once added to the properly normalized tree-level scattering amplitude $\mathcal A_{\rm tree}=Z_{\phi}\times\mathcal A_{\rm tree,0}$. Here the wavefunction renormalization factor $Z_\phi$ needs to be computed from the self-energy correction due to the $\psi\phi^2$ vertex. As in all the previous cases we then find that the $t$-pole cancels leaving only the original pole due to the tree-level graviton exchange
\be
\mathcal A_{t-\text{pole}}=Z_{\phi}\times\mathcal A_{\rm tree,0}+\mathcal A_{\psi\phi^2,1}+\mathcal A_{\psi\phi^2,2}=-\frac{s^2}{\mpl^2t}+\mathcal O(t^0)\,.
\ee

\subsection{Quartic operators}\label{sec:quartic}
\begin{figure}
\centering
\begin{subfigure}{0.35\textwidth}
\centering
\includegraphics[height=1.6in]{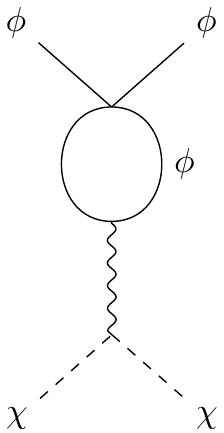}
\caption{One-loop diagram for $\mathcal A_{\phi^4}$.} \label{fig:quartic1}
\end{subfigure}
\hspace*{0.6in} 
\begin{subfigure}{0.35\textwidth}
\centering
\includegraphics[height=1.6in]{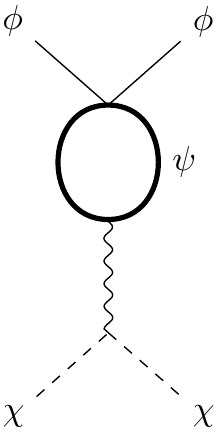}
\caption{One-loop diagram for $\mathcal A_{\psi^2\phi^2}$.} \label{fig:quartic2}
\end{subfigure}
\caption{
The Feynman diagrams for the one-loop scattering processes contributing to $\chi\phi\to\chi\phi$ from the additional quartic EFT operators in \eqref{quartic}. Wiggly lines correspond to graviton propagators, while bold lines correspond to the propagators of the heavy field $\psi$.
} \label{fig:quartic}
\end{figure}

One can also introduce quartic non-derivative interactions between the light field $\phi$ and the heavy field $\psi$ as
\be\label{quartic}
\mathcal L_{(4)}=\sqrt{-g}\left[-\frac{\lambda}{4!}\phi^4 - \frac{\varpi}{4}\psi^2\phi^2\right]\,,
\ee
where $\lambda, \varpi$ are dimensionless couplings. These give new interaction vertices $V_{\phi^4}=-i\lambda$ and $V_{\psi^2\phi^2}=-i\varpi$ allowing for new processes contributing to the $\chi\phi\to\chi\phi$ scattering, shown in Fig.~\ref{fig:quartic}.

\subsubsection{The $\phi^4$ interaction}
The scattering amplitude for the process in Fig.~\ref{fig:quartic1} involves a loop of the light field $\phi$, so we regularize it by adding a mass term $-\frac{1}{2}m^2\phi^2$. The amplitude can then be written as
\be
i\mathcal A_{\phi^4}(s,t)=\frac{1}{\mathcal N}\int\frac{\text{d}^{d}p}{(2\pi)^{d}}V_{\phi^4}\Delta_{\phi}(p)\Delta_{\phi}(k_{1}+k_3-p)V_m^{\mu\nu}D_{\mu\nu;\alpha\beta}(k_{2}+k_{4})V_0^{\alpha\beta}\,,
\ee
where $\mathcal N$ is the symmetry factor of the diagram. After the usual manipulations we find
\be
\mathcal A_{\phi^4}(s,t)=\frac{i\lambda}{\mathcal N\mpl^2}\mu^\epsilon\int_0^1\text{d}x\int\frac{\d^dk}{(2\pi)^d}\frac{\frac{d-2}{d}k^2+m^2}{[k^2+m^2-tx(1-x)]^2}\,.
\ee
This amplitude only depends on $t$ and thus does not contribute to the positivity bounds \eqref{positivity}. Moreover, it has a branch cut for $t\geq 4m^2$, but does not have a $t$-pole. It is thus finite in the forward limit.

\subsubsection{The $\psi^2\phi^2$ interaction}
The quartic $\psi^2\phi^2$ interaction leads to an additional scattering process depicted in Fig.~\ref{fig:quartic2}. Since this process only contains a heavy loop we can again set $m=0$. The scattering amplitude is found to be
\be
i\mathcal A_{\psi^2\phi^2}(s,t)=\int\frac{\text{d}^{d}p}{(2\pi)^{d}}V_{\psi^2\phi^2}\Delta_{\psi}(p)\Delta_{\psi}(k_{1}+k_3-p)V_M^{\mu\nu}D_{\mu\nu;\alpha\beta}(k_{2}+k_{4})V_0^{\alpha\beta}\,,
\ee
giving for the final result
\be
\mathcal A_{\psi^2\phi^2}(s,t)=\frac{i\varpi}{\mpl^2}\mu^\epsilon\int_0^1\text{d} x\int\frac{\text{d}^{d}p}{(2\pi)^{d}}\frac{\frac{d-2}{d}k^2}{[k^2+M^2-tx(1-x)]^2}\,.
\ee
As for the $\phi^4$ vertex, this amplitude is finite at $t=0$ and does not affect the positivity bounds. As expected, it has a branch cut for $t\geq 4M^2$.

\section{Renormalizable spectator field interactions}\label{app:addmore}
\begin{figure}
\centering
\begin{subfigure}{0.3\textwidth}
\centering
\includegraphics[height=1.5in]{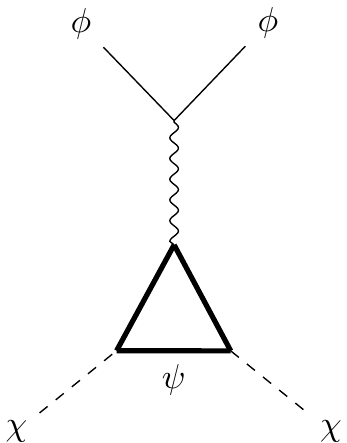}
\caption{One-loop diagram for $\mathcal A_{\chi\psi^2}$} \label{fig:uv1}
\end{subfigure}
\hspace*{0.5in} 
\begin{subfigure}{0.3\textwidth}
\centering
\includegraphics[height=1.5in]{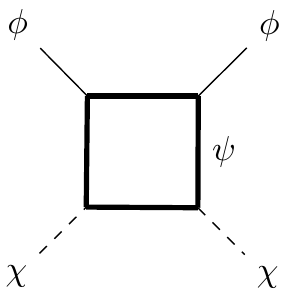}
\caption{One-loop diagram for $\mathcal A_{\rm box}$} \label{fig:uv2}
\end{subfigure}
\caption{
The Feynman diagrams for the one-loop scattering processes contributing to $\chi\phi\to\chi\phi$ arising from the new UV operators in \eqref{pUV_2}. Wiggly lines correspond to graviton propagators, while bold lines correspond to the propagators of the heavy field $\psi$.} \label{fig:newUVapp}
\end{figure}

\begin{figure}[h]
\centering
\begin{subfigure}{0.2\textwidth}
\includegraphics[height=1.5in]{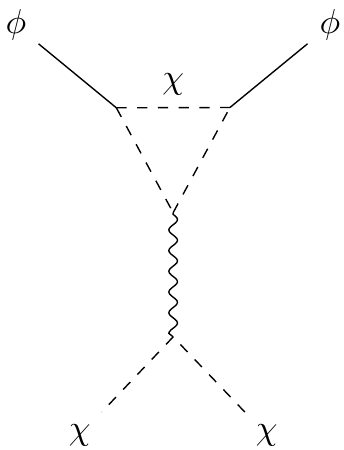}
\end{subfigure}
\hspace*{0.5in} 
\begin{subfigure}{0.2\textwidth}
\includegraphics[height=1.5in]{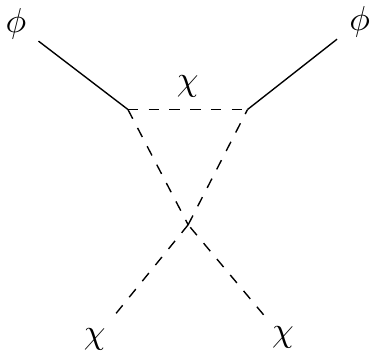}
\end{subfigure}
\caption{Scattering processes contributing to $\chi\phi\to\chi\phi$ at one-loop from the $\phi\chi^2$ interaction. The wiggly line corresponds to the graviton propagator. Note that the diagram on the right also requires a $\chi^4$ self-interaction. } \label{fig:phichi2}
\end{figure}

\begin{figure}[h]
\begin{subfigure}{0.2\textwidth}
\includegraphics[height=1.5in]{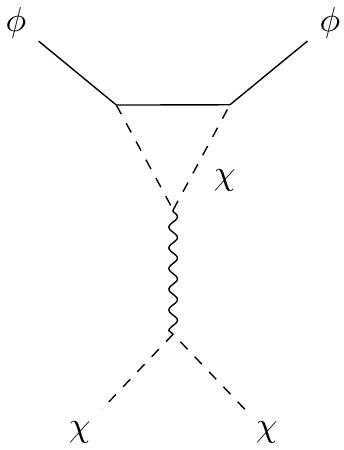}
\end{subfigure}
\hspace*{\fill} 
\begin{subfigure}{0.2\textwidth}
\includegraphics[height=1.5in]{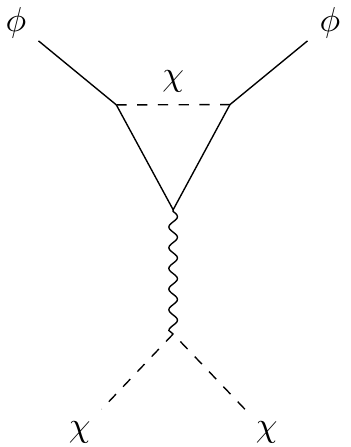}
\end{subfigure}
\hspace*{\fill} 
\begin{subfigure}{0.2\textwidth}
\includegraphics[height=1.5in]{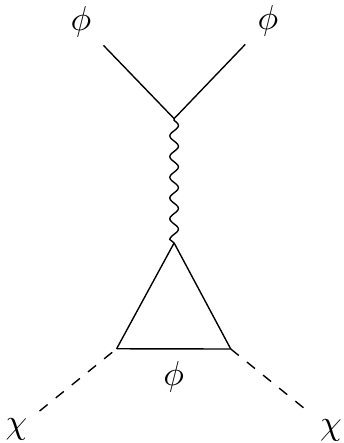}
\end{subfigure}
\hspace*{\fill} 
\begin{subfigure}{0.2\textwidth}
\includegraphics[height=1.5in]{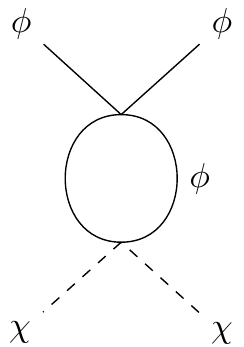}
\end{subfigure}
\caption{Scattering processes contributing to $\chi\phi\to\chi\phi$ at one-loop from the $\chi\phi^2$ interaction. The wiggly lines correspond to the graviton propagators. Note that the last diagram also requires a $\phi^4$ self-interaction.} \label{fig:chiphi2}
\end{figure}

An obvious way of deforming the UV completion in \eqref{pUV_0} would be to introduce non-derivative\footnote{One could in principle also introduce non-derivative interactions directly between the light scalars $\chi$ and $\phi$. The possible cubic operators are $\phi\chi^2$ and $\chi\phi^2$ with the corresponding additional scattering processes shown in Figs.~\ref{fig:phichi2} and~\ref{fig:chiphi2}. The presence of light $\chi$ and $\phi$ loops in some of these diagrams will again require that we introduce a small non-zero masses for $\chi$ and $\phi$ to regulate the amplitude. All these diagrams have been computed earlier and always give negative $s^2$ contribution to the total amplitude. In turn, as was shown in subsection~\ref{sec:quartic} quartic non-derivative interactions do not depend on $s$.} interactions between the spectator $\chi$ and $\psi$. Trying to be as minimalistic as possible let us consider
\be\label{pUV_2}
\L_{\rm pUV,3}=\sqrt{-g}\left[\frac{\mpl^2}{2}R-\frac{1}{2}(\partial\chi)^2-\frac{1}{2}(\partial\phi)^2-\frac{1}{2}(\partial\psi)^2-\frac{1}{2}M^2\psi^2-\alpha M\phi\psi^2-\varrho M\chi\psi^2\right]\,,
\ee
where we have only added the interaction $-\varrho M\chi\psi^2$ leading to a new vertex $V_{\chi\psi^2}=-2i\varrho M$. This introduces two new diagrams contributing to the $\chi\phi\to\chi\phi$ scattering shown in Figs.~\ref{fig:uv1} and \ref{fig:uv2}. Note that as in \eqref{toy} this new interaction would also introduce new terms like $\frac{\varrho^2}{M^2}R^{\mu\nu}\partial_\mu\chi\partial_\nu\chi$ in the IR theory. We shall not work out the IR action explicitly here since it is equivalent to working at the level of scattering amplitudes.

\begin{itemize}
\item The first diagram in Fig.~\ref{fig:uv1} is entirely analogous to the diagram shown in Fig.~\ref{fig:1b} and the amplitude reads
\be
\begin{split}
i\mathcal A_{\chi\psi^2}(s,t)=\int \frac{\d^dp}{(2\pi)^d}&V_{\chi\psi^2}^2 \Delta_{\psi}(p)\Delta_\psi(p+k_2)\Delta_{\psi}(p-k_4)\\
&V_M^{\mu\nu}(p+k_2,k_4-p)D_{\mu\nu;\alpha\beta}(k_1+k_3)V_0^{\alpha\beta}(k_1,k_3)\,.
\end{split}
\ee
Although slightly different in details the final result for the finite $s^2$ contribution gives again the second equality in \eqref{loops2} with $\alpha\leftrightarrow\varrho$ thus leading to  the same negative contribution
\be
\mathcal A_{\chi\psi^2,\,s^2}(s,t)=-\frac{\varrho^2s^2}{90(4\pi)^2M^2\mpl^2}+\dots\,.
\ee

\item The box diagram in Fig.~\ref{fig:uv2} is new. Its amplitude can be computed as
\be
\begin{split}
i\mathcal A_{\rm box}(s,t,u)=&\int\frac{\d^dp}{(2\pi)^d}V_{\phi\psi^2}^2V_{\chi\psi^2}^2\Delta_\psi(p)\Delta_\psi(p+k_3)\Delta_\psi(p+k_3+k_4)\Delta_\psi(p-k_1)\\
&+\text{ crossed diagrams}\,,
\end{split}
\ee
where one also has to add the contributions from the crossed diagrams, similarly as was done in appendix \ref{app:box}.
For the specific diagram in Fig.~\ref{fig:uv2}, the various denominators can be combined by introducing three Feynman parameters $x,y,z$ leading to
\be
i\mathcal A_{\rm box}(s,t)=\frac{16\alpha^2\varrho^2 M^4}{4}\int\frac{\d^dp}{(2\pi)^d}\int_{\mathcal R}\frac{3!\text{d}x\,\text{d}y\,\text{d}z}{[k^2+\Delta]^4}\,,\nonumber
\ee
where $\Delta\equiv M^2+sz(x+y+z-1)-txy$ and the factor $1/4$ appears due to the symmetry of the diagram; the integration region is $\mathcal R= \{x,y,z\in\mathbb R\,|\,x,y,z\geq0, x+y+z\leq 1\}$. The momentum integral can be easily taken using \eqref{dimreg} leaving at $t=0$
\be
\mathcal A_{\rm box}(s,t)=\frac{3\alpha^2\varrho^2M^2}{4\pi^2}\int_{\mathcal R}\text{d}x\,\text{d}y\,\text{d}z\,z^2(-1+x+y+z)^2\times s^2+\mathcal O(t^0)\,.
\ee
Since in the integration region $\mathcal R$ the integrand is positive, so is the integral (and equal to $1/1260$) and so is  the contribution to the $s^2$ term in the scattering amplitude.  Hence, the positive contribution of the box diagram could naively be used to cancel the negative $s^2$ contributions coming from the processes shown in Figs.~~\ref{fig:1b} and \ref{fig:uv1}. However, the situation here is in fact very similar to what we have seen earlier in section \ref{sec:qed} when discussing the positivity bounds \eqref{bound1} obtained from the UV theory in \eqref{scalar_photon2}. Also here the positive contribution coming from the loop diagram in Fig.~\ref{fig:uv2} can be subtracted by the procedure of improved positivity bounds.
\end{itemize}
We thus conclude that none of the possible renormalizable interactions of the spectator field satisfy the new positivity bounds.

\bibliographystyle{JHEP}
\bibliography{references}

\end{document}